\documentclass[10pt,twocolumn]{article}

\usepackage[utf8]{inputenc}
\usepackage[T1]{fontenc}
\usepackage[a4paper,margin=1.8cm]{geometry}
\usepackage{graphicx}
\usepackage{amsmath,amssymb}
\usepackage{siunitx}
\usepackage{booktabs}
\usepackage{authblk}
\usepackage{titlesec}
\usepackage{abstract}
\usepackage{eurosym}
\usepackage{makecell}

\usepackage{booktabs}
\usepackage{tabularx}
\usepackage{array}

\usepackage{hyperref}
\hypersetup{
  colorlinks=true,
  linkcolor=blue,
  citecolor=blue,
  urlcolor=blue
}

\titleformat{\section}{\large\bfseries}{\thesection.}{0.5em}{}
\titleformat{\subsection}{\normalsize\bfseries}{\thesubsection.}{0.5em}{}

\newcolumntype{L}[1]{>{\raggedright\arraybackslash}p{#1}}
\newcolumntype{C}[1]{>{\centering\arraybackslash}p{#1}}

\begin{document}

\title{\bfseries On the velocity range limitation of DBSCAN for event-based particle tracking and a proposed solution}
\author[1]{Nicolas Lanchon}
\author[2]{Charlotte Vienney}
\author[2]{Samuel Bera}
\author[2]{Cheng Wang}
\author[1]{Adam Cheminet}
\author[2]{Nicolas Plihon}
\author[2]{Romain Volk}
\author[2]{Mickaël Bourgoin}
\author[1]{Bérengère Dubrulle}
\affil[1]{Université Paris-Saclay, CEA, CNRS, SPEC,Gif-sur-Yvette 91191, France}
\affil[2]{Université de Lyon, ENS de Lyon, CNRS, Laboratoire de physique, Lyon 69342, France}
\date{}

\twocolumn[
  \maketitle
  \begin{onecolabstract}
    Event-based cameras offer new opportunities for particle tracking in fluid mechanics by providing asynchronous, sparse measurements with high temporal resolution and no fixed acquisition frequency. Processing such data requires algorithms capable of fully exploiting these distinctive features. A natural first approach to perform particle tracking on event-based data is to apply a clustering algorithm, in particular DBSCAN, as it does not require specifying the number of clusters a priori. In this paper, we show, using a simplified synthetic two-particle system, that this classical algorithm is severely limited in terms of the accessible velocity range $v_{\max}/v_{\min}$. To address this limitation, and to take advantage of the absence of a fixed acquisition frequency in event-based cameras, we propose a new two-dimensional, two-component (2D2C) particle-tracking algorithm based on clustering. The method relies on a recursive application of DBSCAN with multiple temporal normalizations, retaining only clusters that satisfy a set of validation criteria. A key feature of the proposed algorithm is that it does not assume any characteristic time scale or typical velocity. The algorithm is evaluated on realistic synthetic datasets generated from numerical simulations of turbulent flows, and the influence of key parameters is systematically investigated. Particular attention is given to the definition of evaluation metrics adapted to clusters of events for particle tracking. Finally, the proposed algorithm is applied to experimental data acquired in a turbulent flow generated by a von Kármán apparatus, demonstrating its ability to extract particle trajectories from real event-based measurements.

    \vspace{0.3cm}    
    This is a preprint of an article published in Experiments in Fluids. The final authenticated version is available online at: \href{https://doi.org/10.1007/s00348-026-04313-z}{https://doi.org/10.1007/s00348-026-04313-z}.
  \end{onecolabstract}
  \vspace{1em}
]


\section{Introduction}

One of the essential features of turbulent flows is their multi-scale nature, arising from the multitude of vortices that compose them. Accurately measuring such a wide range of scales with a camera is technically challenging, notably because conventional cameras generally sample the flow using a single characteristic timescale (or two in the case of double-frame cameras), determined by their acquisition frequency. As a result, the range of measurable timescales is limited: at the lower end by the acquisition frequency, and at the upper end by the camera’s maximum memory capacity. This limitation can be particularly significant for high-Reynolds-number flow visualization and may compromise the accuracy of turbulent velocity-field measurements, leading to effects such as gradient smoothing, energy-spectrum truncation, and increased uncertainty at low velocities. Moreover, some turbulent flows exhibit dynamics that are much slower than those underlying the turbulence itself. Examples include turbulent bifurcations in von Kármán flows~\cite{Cortet2010,Saint-Michel2013,Dubrulle2019}, the development of certain boundary layers~\cite{Obligado2013,Gayout2021}, or flows belonging to a regime of weakly nonlinear wave turbulence~\cite{Lanchon2023,Lanchon2025}. Studying such flows requires both high acquisition frequencies, to properly resolve the fastest dynamics, and very long acquisition times, to capture the slow dynamics. This generates prohibitive data volumes that complicate their analysis, or even make it impossible.

\begin{table*}
  \centering
\label{tab:biblio}
    \begin{tabular}{
    L{0.19\textwidth}
    L{0.42\textwidth}
    C{0.20\textwidth}
    C{0.08\textwidth}
  }
  \toprule
  Approach & Reference & Primary computed object & Dim. \\
  \midrule
  Image reconstruction 
  & Howell \textit{et al.} (2020), \cite{Howell2020} 
  & Particle tracking & 2D2C \\
  & Willert \textit{et al.} (2022) \cite{Willert2022} , \textit{sum-of-correlation} 
  & Eulerian field & 2D2C \\
  & Willert (2023) \cite{Willert2023} 
  & Eulerian field & 2D2C \\
  & Willert \textit{et al.} (2025) \cite{Willert2025} 
  & Particle tracking & 3D3C \\
  & Franceschelli \textit{et al.} (2025) \cite{Franceschelli2025} 
  & Eulerian field & 2D2C \\
  & Sachs \textit{et al.} (2026) \cite{Sachs2026} 
  & Particle tracking & 2D2C \\
  \midrule
  Kalman-like-filter
  & Drazen \textit{et al.} (2011) \cite{Drazen2011} 
  & Particle tracking & 2D2C \\
  & Borer \textit{et al.} (2017) \cite{Borer2017} 
  & Particle tracking & 3D3C \\
  & Rusch \textit{et al.} (2023) \cite{Rusch2023}
  & Particle tracking & 3D3C \\
  & AlSattam \textit{et al.} (2024) \cite{AlSattam2024} 
  & Particle tracking & 2D2C \\
  \midrule
  Clustering-based 
  & Hu \textit{et al.} (2022) \cite{Hu2022} 
  & Object tracking & 2D2C \\
  & Bera \textit{et al.} (2026) \cite{Bera2026}
  & Particle tracking & 2D3C \\
  & Sachs \textit{et al.} (2026) \cite{Sachs2026} 
  & Particle tracking & 2D2C \\
  & Present paper 
  & Particle tracking & 2D2C \\
  \midrule
  Other approaches 
  & Ni \textit{et al.} (2012) \cite{Ni2012} 
  & Brownian motion tracking & 2D2C \\
  & Wang \textit{et al.} (2020) \cite{Wang2020} 
  & Particle tracking using optical flow & 3D3C \\
  & Willert \textit{et al.} (2022)~\cite{Willert2022} , \textit{motion~compensation}
  & Eulerian field & 2D2C \\
  \bottomrule
  \end{tabular}
    \caption{Classification of algorithms proposed in the literature for flow visualisation using event-based data as well as its dimensionality. It is worth noting that in some papers, once the primary object is computed, authors compute from it others observables such as the Eulerian field (at 2D or 3D), shear stress, etc. The dimensionality reported here refers to the final physical quantities targeted by each algorithm. Comparisons between several methods are provided in Refs.~\cite{AlSattam2024, Gunasekaran2025}.}
\end{table*}

Recently, a promising new camera technology that could change the game has emerged: the event-based camera~\cite{Gallego2020}, also referred to as dynamic vision sensor (DVS). Instead of measuring light intensity directly, such cameras detect changes in light intensity and output a stream of events indicating when and where sufficiently large variations occur. Each event is assigned a polarity, positive or negative, depending on whether it corresponds to an increase or a decrease in light intensity. Formally, if $I(t_k)$ denotes the light intensity at the pixel at the instant when the $k$-th event is generated, a new event is triggered when the intensity variation satisfies

\begin{equation}
\label{eq:thresholdSimple}
I(t_{k+1}) - I(t_{k}) = 
    \left\{\begin{array}{cc}
          c_+ I(t_k) & \text{for a positive event}\, , \\
         -c_- I(t_k) & \text{for a negative event}\, ,
    \end{array} \right.
\end{equation}
where $c_+$ and $c_-$ are relative intensity thresholds. It is often more convenient to express this condition in terms of the logarithm of the light intensity, $L = \log I$, leading to
\begin{equation}
\label{eq:thresholdLog}
L(t_{k+1}) - L(t_{k}) = 
    \left\{\begin{array}{ll}
          C_+  & \text{for a positive event} \, , \\
         -C_- & \text{for a negative event} \, ,
    \end{array} \right.
\end{equation}
with $C_+ = \log(1+c_+)$ and $C_- = - \log(1-c_-)$. 
For commercial event-based cameras, typical values of threshold $c_+$ and $c_-$ range from $10$ to $50$~\% of the light intensity~\cite{Gallego2020}. In the special case where $C = C_+ = C_-$, the signal returned by the event-based camera can be written analytically as
\begin{equation}
    \left\lfloor \frac{L(t+\delta t)}{C} \right\rfloor - \left\lfloor \frac{L(t)}{C} \right\rfloor \, ,
\end{equation}
where $\delta t$ denotes the time step of the camera internal clock and $\lfloor . \rfloor$ the floor function. For each pixel, this quantity is equal to $+1$ if a positive event occurs at time $t$, $-1$ if a negative event occurs, and $0$ otherwise.

In addition, event-based cameras usually include configurable high-pass and low-pass filters applied to the light intensity signal prior to event generation. The exact behavior of these filters is complex and remains the subject of ongoing research. In some configurations, it is possible to suppress the acquisition of negative events while preserving positive events, without requiring the light intensity to diverge. The low-pass filter can be approximately modeled as a causal first-order filter whose cutoff frequency increases with the logarithm of the light intensity~\cite{McReynolds2022,Graca2021}.

Event-based camera technology offers several advantages for flow visualization in particle-seeded fluids~\cite{Willert2022}. First, because these cameras, by construction, only record changes in light intensity, the resulting data stream is much lighter than that produced by a conventional camera. In particular, the background is not recorded. The data have a sparse structure, which enables longer recording durations and significantly reduces computational requirements, as well as camera-to-computer transfer times. Second, due to their asynchronous nature, event-based cameras do not have a fixed acquisition frequency and therefore provide a truly multi-timescale measurements. They can reliably capture, during the same acquisition, particle motion over velocity ranges spanning several orders of magnitude. Third, the overall recording system is relatively affordable. Indeed, owing to their high sensitivity, event-based cameras do not require very powerful illumination. In practice, a continuous diode laser costing only a few hundred euros is sufficient for 2D2C flow visualization~\cite{Willert2022}. In addition, no synchronization unit between the camera and the laser is required.

However, for the sake of completeness, it is worth noting that event-based cameras also have limitations that may be addressed in future generations. These include a relatively low pixel count compared to state-of-the-art frame-based sensors and data binarization, which results in a loss of information regarding the particle sub-pixel position. Additional limitations include pixel latency~\cite{Lichtsteiner2008}, which introduces uncertainty in the event timestamp (typically between $0.1$ and $1$~ms), and a maximum event rate beyond which the sensor saturates, thereby limiting the maximum usable particle density. It is also worth noting that event-based cameras exhibit uncertainty in the threshold coefficients $C_+$ and $C_-$, which can reach up to 10\% of their values~\cite{McReynolds2022,Lichtsteiner2008,Posch2011}.

Since the 2010s—and even more so since the early 2020s, following the release of commercial event-based cameras with sufficiently high spatial resolution—many research groups have investigated flow visualization using these devices. From an algorithmic perspective, existing approaches can be classified into four main categories (see Tab.~\ref{tab:biblio}). The first approach consists of reconstructing images, often referred to as pseudo-images, from event-based data by accumulating positive-polarity events over a given time window~\cite{Willert2022, Howell2020, Willert2023, Willert2025, Franceschelli2025}. After applying a Gaussian filter to mitigate peak-locking effects, a sequence of images is obtained, on which classical frame-based algorithms can be applied. Although this approach largely ignores the specific characteristics and novel capabilities of event-based cameras, it enables relatively inexpensive experimental setups due to the low illumination levels required by these sensors. However, the binarized nature of the data and the lack of precise light-intensity information hinder frame-based algorithms and limit their performance. A second approach relies on Kalman filtering (or Kalman-like-filter algorithms)~\cite{Drazen2011,Borer2017,Rusch2023,AlSattam2024}. In this case, the dataset is processed on an event-by-event basis. Each event is associated with an existing track, or creates a new one, according to various validation criteria, and the particle position corresponding to that track is updated using a linear combination of the previous estimated position and the new event location. Kalman-filter-based approaches are particularly well suited for real-time processing~\cite{Drazen2011,Rusch2023}. However, when the data are processed offline, the inherent causality of such filters appears to impose an unnecessary constraint and, in particular, can lead to poor estimates near track initialization. The third approach, adopted in the present study, is based on clustering algorithms~\cite{Hu2022,Bera2026, Sachs2026}, and more specifically on variants of DBSCAN, a clustering algorithm that does not require the number of clusters to be specified in advance. Rather than processing the data event by event, this approach aims to process the dataset globally and to identify entire trajectories. Finally, several other approaches have been reported in the literature. These include particle tracking for Brownian motion with a particular focus on sub-pixel localization~\cite{Ni2012}, the use of optical-flow algorithms to track particles~\cite{Wang2020}, and the reconstruction of Eulerian velocity fields using a variational principle within a spatio-temporal subdomain of the data (\textit{motion compensation} in \cite{Willert2022}).

\begin{figure}[h]
    \centering
    \includegraphics[width=7cm]{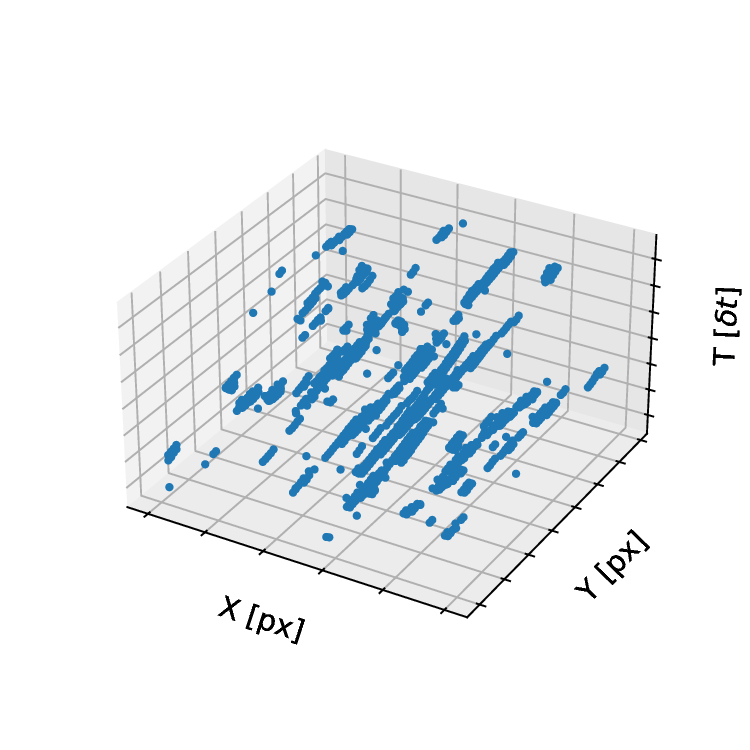}
    \caption{Spatio-temporal representation of a subdomain of the experimental data presented in Sec.~\ref{sec:Expe}.}
    \label{fig:introExp}
\end{figure}

As previously stated, clustering algorithms appear particularly well suited for data of illuminated moving particles from event cameras. Indeed, by examining the spatio-temporal representation of the data (see, for instance, Fig.~\ref{fig:introExp}), one can easily see particle trajectories 'at first glance'. Each trajectory appears as a cluster of elongated spatiotemporal points. It is therefore natural to consider clustering algorithms, as they are well adapted to sparse datasets, and their effectiveness is further enhanced by the fact that events generated by particle are generally continuous in space. Once trajectories have been identified as clusters of events, they can be fitted to extract positions, velocities, and accelerations (bearing in mind that events indicate the position of the particle boundary rather than its center). In this study, we will mainly discuss the first step : the event clustering.


In the present study, we explore the use of the DBSCAN algorithm to detect trajectories within event-based datasets. The choice of DBSCAN is motivated by the fact that it does not require the number of expected clusters (i.e., here, the number of expected trajectories) to be specified a priori. After a brief description of how DBSCAN can be applied to spatiotemporal event-based data (Sec.~\ref{sec:clusterDBSCAN}), we investigate its behavior using a simple system composed of two parallel trajectories (Sec.~\ref{sec:understandingDBSCAN}). We show that, even in this simplified configuration, DBSCAN is severely limited in terms of the velocity range $v_{\max}/v_{\min}$ over which trajectories can be correctly detected. To address this limitation, we develop a new 2D2C particle-tracking algorithm capable of detecting trajectories over an arbitrarily wide velocity range (Sec.~\ref{sec:clusterAlgo}). The performance of this algorithm is then evaluated on synthetic datasets using appropriate evaluation metrics to quantify the quality of trajectory detection through clustering (Sec.~\ref{sec:multipleParticle}). Finally, we apply the proposed clustering method to experimental data from a von Kármán turbulent flow (Sec.~\ref{sec:Expe}).

\section{Event Clustering Using DBSCAN}
\label{sec:clusterDBSCAN}

To extract particle positions from raw event data, our initial approach was to apply the classical DBSCAN algorithm~\cite{Ester1996,Schubert2017}. Indeed, unlike many other clustering algorithms, DBSCAN does not require the number of clusters to be specified in advance, an important advantage in our context, since the number of particles within the observation field is unknown and may vary over time. From a user’s perspective, DBSCAN requires two parameters: a neighborhood radius $\varepsilon$ and a minimum number of events required to form a dense neighborhood, denoted \textit{minSamples} following the \textit{scikit-learn} documentation. The algorithm identifies events that have at least \textit{minSamples} neighbors within their $\varepsilon$-neighborhood and designates them as \textit{core points}. Any event lying within the $\varepsilon$-neighborhood of a core point is then assigned to the same cluster as it. Events with fewer than \textit{minSamples} neighbors and not within the neighborhood of any core point are classified as noise and discarded. In this study, we use the Python implementation of DBSCAN provided by the \textit{scikit-learn} library.

The main challenge in applying DBSCAN to our data lies in the fact that events are defined in a spatiotemporal domain, with two spatial coordinates (in pixels) and one temporal coordinate (typically in µs). To measure distances in this space, we introduce the following Euclidean-like metric that combines spatial and temporal dimensions:
\begin{equation}
d(e_i, e_j) = \sqrt{(x_i - x_j)^2 + (y_i - y_j)^2 + \alpha^2 (t_i - t_j)^2} \, ,
\label{eq:metriqueAlpha}
\end{equation}
where $\alpha$ is a normalization time parameter in px/µs. This parameter, which can be interpreted as a characteristic velocity, plays a central role throughout this paper.

For DBSCAN to correctly detect trajectories, the parameter $\alpha$ must be carefully tuned. If $\alpha$ is too small, the time axis is compressed, and two trajectories passing through the same pixel at different times may be merged into the same cluster. Conversely, if $\alpha$ is too large, the time axis is expanded, and events belonging to the same trajectory may become too distant from one another to fall within the same $\varepsilon$-neighborhood, leading to fragmented trajectories across multiple clusters or even their rejection by DBSCAN. Besides, since by definition turbulent flows contain a wide range of scales, the choice of $\alpha$ is necessarily an arbitrary trade-off which will ultimately limits the range of time scales (and thus the range of velocities) that can be accurately detected.

\section{Analysis of DBSCAN clustering
from a simple system}
\label{sec:understandingDBSCAN}

To investigate the behavior of DBSCAN on events generated by particle trajectories, we first simplify the problem by considering a toy model consisting of two particles with specified velocities and separation distances.

\subsection{Generation of synthetic data}
\label{sec:genereSyntheData}

Synthetic datasets used for evaluation are generated with the following procedure. We first assume that each particle produces a light intensity distribution on the sensor, also referred to as the point-spread function~\cite{Champagnat2014}, of the form
\begin{equation}
    \begin{split}
    h(x,y) =  \frac{1}{4} &\left[ \text{erf}\left( \frac{x+1/2}{\sqrt{2} \sigma_p}\right) - \text{erf}\left( \frac{x-1/2}{\sqrt{2} \sigma_p} \right) \right] \times \\
    &\left[ \text{erf}\left( \frac{y+1/2}{\sqrt{2} \sigma_p}\right) - \text{erf}\left( \frac{y-1/2}{\sqrt{2} \sigma_p} \right) \right]\,,
    \end{split}
\end{equation}
where $\sigma_p$ denotes the standard deviation of the point-spread function, hereafter referred to as the particle size, and $\text{erf}$ is the error function. This intensity profile corresponds to the local integration, over a each pixel, of a Gaussian light pulse scattered by the particle. The corresponding particle image size associated with this point-spread function is therefore typically about $6\sigma_p$. However, in the context of event-based data, the notion of ``event footprint width'', i.e., the width of the trajectory expressed in terms of the number of events, appears to be more relevant. To model the event-based camera, a synthetic image of the light intensity is generated at each time step $\delta t$ ---mimicing
the camera’s internal clock--- by summing the contributions of all particles such as
\begin{equation}
I(\mathbf{x},t) = \sum_{p=1}^{N} I_p h(\mathbf{x}-\mathbf{x}_p(t)) \, ,
\end{equation}
where $\mathbf{x} = (x,y)$, $N$ is the number of particles, and $I_p$ and $\mathbf{x}_p$ denote the nominal light intensity and the position of particle $p$, respectively. The value of $I_p$ is taken to be identical for all particles, and a minimum light intensity $I_{\min}$ is imposed in order to define a signal-to-noise ratio $I_{\max}/I_{\min} =  \text{erf}(\sqrt{2}/\sigma_p)I_p /I_{\min} $. It is worth mentioning that some authors add noise to $I_0$ to take into account defocusing~\cite{Sachs2026}. For each pixel, an event is generated when the variation of the logarithm of the light intensity verifies Eq.~(\ref{eq:thresholdLog}). To simplify, we restrict ourselves to the case of symmetric logarithmic thresholds, $C_+ = C_- = C$, and do not include any noise on them. To determine the first event at each pixel, the logarithm of the light intensity $L$ is compared to $\log I_{\min}$. To ensure physical consistency, the sampling time step $\delta t$ must be sufficiently small, and particle velocities must remain much smaller than $1$~px$/\delta t$. For reference, the internal time step of the event-based camera used in Sec.~\ref{sec:Expe} is $\delta t = 1$~µs, and particle velocities in our experimental setup are typically smaller than $10^{-3}$~px$/\delta t$ (i.e. $1000$~px/s). Finally, in this paper, we process only positive-polarity events.

\subsection{The two particles test}
\label{sec:twoPartTest}

\begin{figure}[h]
    \centering
    \includegraphics[width=5cm]{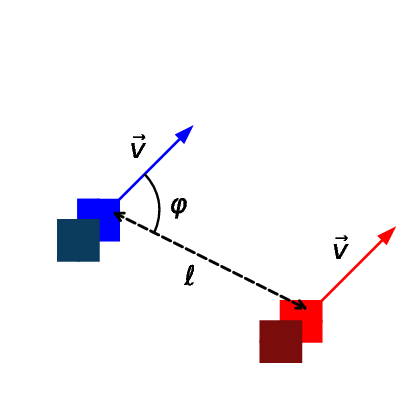}
    \caption{Schematic view of the event footprints produced by two idealized particles separated by a distance $\ell$ and moving at a velocity $v = \vert \vert \vec{v } \vert \vert$. Negative events are shown here for illustration (in dark blue and dark red) but are not used in the processing.}
    \label{fig:schema2particules}
\end{figure}

In this section, we evaluate the DBSCAN algorithm with a simplified toy-model system composed of two particles moving at the same velocity $\vec{v}$ and separated by a distance $\ell$. The angle between the particle pair and the velocity vector is denoted by $\varphi$ (see Fig.~\ref{fig:schema2particules}). We then apply the DBSCAN algorithm on the positive-polarity event generated in this configuration, examine the results obtained and analyze its sensitivity to various parameters.

\begin{figure*}[h]
    \centering
    \includegraphics[width=16cm]{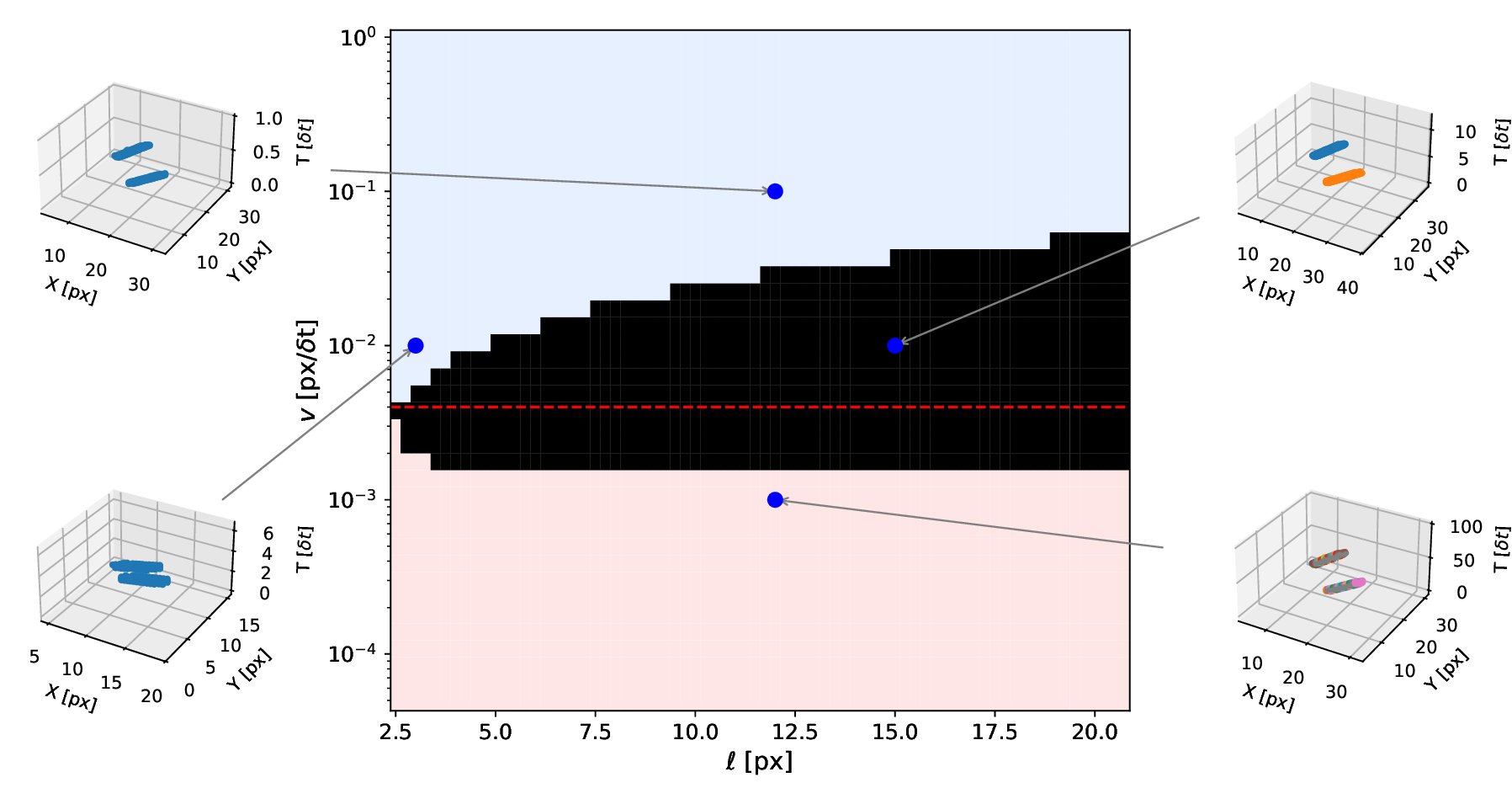}
    \caption{Phase diagram indicating the success or failure of particle clustering for velocity $v$, separation distance $\ell$, and $\varphi = 0$. The parameters used in the simulations and in the DBSCAN algorithm are listed in Tab.~\ref{tab:parameters2Particles}. The black region corresponds to successful clustering, while the light red and light blue regions correspond to failures. The red dashed line marks $v = \alpha$. Surrounding panels show spatio-temporal plots of the data for four representative pairs $(\ell,v)$, with one color per detected cluster and gray points representing noise. Time units are expressed in terms of $\delta t$, i.e., the period of the internal camera clock.}
    \label{fig:phaseDiagram}
\end{figure*}

In a first test, Figure~\ref{fig:phaseDiagram} presents the results of the DBSCAN algorithm for a range of particle velocities and separations in the case where one particle follows the other ($\varphi = 0$). The parameters used in these simulations are listed in Tab.~\ref{tab:parameters2Particles}. This figure can be interpreted as a phase diagram indicating, for a given set of parameters, the pairs $(v,\ell)$ for which the clustering succeeds (i.e., when the algorithm correctly identifies two separate particles, shown in black) or fails (shown in light blue and light red). We observe that the algorithm successfully distinguishes particles separated by as little as $2.5$~px. Considering that the particle event footprint has a half-moon shape approximately $3$~px wide, and that the neighborhood radius used by DBSCAN is $\varepsilon = 1.5$~px, an estimation of the minimum distance for successful separation is about $3/2 + 1.5 = 3$~px. The minimum separation observed in Fig.~\ref{fig:phaseDiagram} therefore appears to be similar to the theoretical optimum. Along the vertical axis, the results show that particles are correctly clustered when their velocities are comparable to $\alpha$, within approximately one order of magnitude. At $\ell = 10$~px, the range of accessible velocity is $v_{\max}/v_{\min} \approx 14$. 

Interestingly, the phase diagram reveals two distinct types of failure. The first one, shown in light blue in Fig.~\ref{fig:phaseDiagram}, occurs when the neighborhood radius $\varepsilon$ exceeds the minimum spatio-temporal distance between the footprint of the two particle trajectories. In this case, DBSCAN merges the two sets of events and identifies them as a single large cluster. The second failure mode appears at low velocities, when the events become highly separated in time (see the vertical axis in the bottom-right inset). Here, the number of events within the neighborhood of each point falls below \textit{minSamples}, the minimum required to define a core point, causing the trajectory to fragment into multiple clusters or be classified as noise. This velocity threshold does not depend on $\ell$, since it concerns each particle independently, but it is strongly influenced by experimental parameters such as the signal-to-noise ratio $I_{\max}/I_{\min}$, the event-based threshold $C$, the particle size $\sigma_p$, and the DBSCAN parameters (see below). 

\begin{table*}[h]
  \centering
  \begin{tabularx}{\textwidth}
  {>{\centering\arraybackslash}X
   >{\centering\arraybackslash}X|
   >{\centering\arraybackslash}X
   >{\centering\arraybackslash}X|
   >{\centering\arraybackslash}X
   >{\centering\arraybackslash}X}
  \toprule
  Simulation parameters & Value & Event camera settings & Value & Clustering parameters & Value \\
  \midrule
  $\varphi$ & $0^\circ$ & $C$ & 0.4& $\alpha$ & 0.004 px/$\delta t$  \\
  $\sigma_p$ & 0.6 px & $c_+$ & 49~\% & $\varepsilon$ & 1.5 px  \\
  $I_{\max}/I_{\min}$ & 10 & -$c_-$ & -33~\% & \textit{minSamples} & 6  \\
  \bottomrule
  \end{tabularx}
  \caption{Parameters used in the simulations to produce the phase diagram in Fig.~\ref{fig:phaseDiagram} and panel (a) of Fig.~\ref{fig:phaseDiagralVariousParameters}. Here, $\varphi$ is the angle between the particle pair and their velocity vector (see Fig.~\ref{fig:schema2particules}), $\sigma_p$ is the particle size, $I_{\max}/I_{\min}$ is the signal-to-noise ratio of the light intensity $I$, $\alpha$ is the temporal normalization factor defined in Eq.~(\ref{eq:metriqueAlpha}), $\varepsilon$ is the neighborhood radius used by DBSCAN, \textit{minSamples} is the minimum number of points required to form a dense neighborhood in DBSCAN, and $C$ is the logarithmic threshold used to define an event according to Eq.~(\ref{eq:thresholdLog}). For reference, $c_+$ and $c_-$ denote the corresponding thresholds on $I$ defined by Eq.~(\ref{eq:thresholdSimple}). }
  \label{tab:parameters2Particles}
\end{table*}

\begin{figure*}[h!]
    \centering
    \includegraphics[width=16cm]{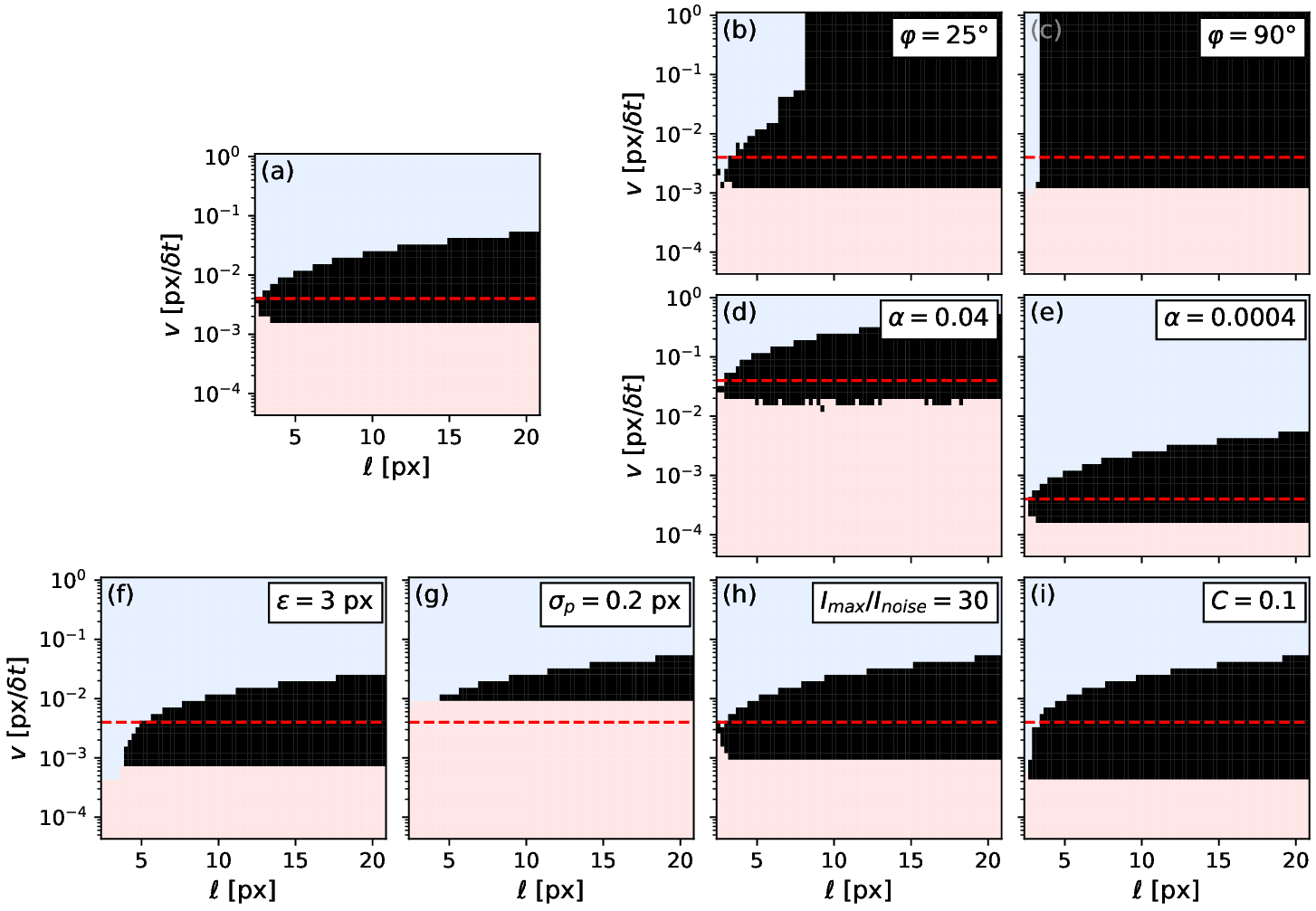}
    \caption{Influence of various parameters on the phase diagram. Panel (a) corresponds to the reference case shown in Fig.~\ref{fig:phaseDiagram}, with parameters listed in Table~\ref{tab:parameters2Particles}. Each of the other panels was computed using the same parameters, except for the one indicated in the legend.}
    \label{fig:phaseDiagralVariousParameters}
\end{figure*}
To better understand how the phase diagram is affected by various parameters, Fig.~\ref{fig:phaseDiagralVariousParameters} presents several phase diagrams in which, from panels (b) to (i), only one parameter has been varied relative to its reference value given in Table~\ref{tab:parameters2Particles}. Panel (a) corresponds to the reference case, which is the same as that shown in Fig.~\ref{fig:phaseDiagram}.
\begin{itemize}
    \item[•] \textbf{Angle between the particles and their velocity ($\varphi$).} Panels (a), (b), and (c) show the phase diagrams for $\varphi = 0^\circ$, $25^\circ$, and $90^\circ$, respectively. While the lower velocity limit remains unaffected by $\varphi$, the upper limit changes significantly. For $\varphi \neq 0^\circ$, a critical inter-particle distance appears, beyond which there is no upper velocity limit for successful clustering. This critical distance decreases as $\varphi$ increases, and for $\varphi = 90^\circ$, no upper bound on velocity is observed. In this configuration, the algorithm becomes fully multi-time-scale. However, in real datasets, all angles $\varphi$ are possible. Therefore, in this section, we discuss the most restrictive case ($\varphi = 0^\circ$), keeping in mind that the algorithm can perform better for other orientations.
    \item[•] \textbf{Time normalization ($\alpha$).} Panels (a), (d), and (e) show the results of the DBSCAN algorithm for $\alpha = 4 \times 10^{-3}$, $4 \times 10^{-2}$, and $4 \times 10^{-4}$~px/$\delta t$, respectively. Varying $\alpha$ does not modify the shape of the success domain but rather shifts it along the velocity axis. The value of $\alpha$ thus corresponds to the characteristic velocity at which particles are correctly separated by the algorithm. This property forms the cornerstone of the extended algorithm described in Section~\ref{sec:notreAlgo}.
    \item[•] \textbf{Neighborhood radius ($\varepsilon$).} Panels (a) and (f) compare the results for $\varepsilon = 1.5$ and $3$~px. Increasing $\varepsilon$ slightly changes the shape of the success domain and allows correct clustering for lower velocities relative to $\alpha$. More importantly, a larger $\varepsilon$ increases the minimum separation distance required for successful clustering, from $2.5$~px at $\varepsilon = 3$~px to $4$~px at $\varepsilon = 3$~px. To handle higher particle densities, $\varepsilon$ should therefore be minimized while keeping it greater than $1$.
    \item[•] \textbf{Particle size ($\sigma_p$).} Panels (a) and (g) compare results for $\sigma_p = 0.6$ and $0.2$~px, corresponding to event footprints (i.e., the set of events generated by a particle) of $3$ and $1$ events wide, respectively. When the event footprint is reduced to a single event, significantly fewer neighboring events are available for each event, causing trajectories to be fragmented more easily. As a consequence, the accessible velocity range decreases dramatically for smaller particles.
    \item[•] \textbf{Signal-to-noise ratio ($I_{\max}/I_{\min}$).}  Panels (a) and (h) compare simulations with signal-to-noise ratios of 10 and 30, respectively. A higher $I_{\max}/I_{\min}$ increases the velocity range over which clustering is successful. This effect can be explained by the larger number of events generated per particle, which reduces the temporal separation between successive events and therefore mitigates trajectory fragmentation.
    \item[•] \textbf{Threshold ($C$).} Similarly to the signal-to-noise ratio, comparing panels (a) and (i) shows that lowering the threshold $C$ increases the number of generated events, thereby reducing trajectory fragmentation and extending the accessible velocity range.
\end{itemize}

To summarize, clustering performance improves when the particle alignment is more perpendicular to their velocity, $\alpha$ is related to the characteristic velocity of successful clustering, and the neighborhood radius $\varepsilon$ has to be small. Several experimental parameters, such as $\sigma_p$, $C$, and $I_{\max}/I_{\min}$, strongly influence the event footprint and density, and thus the efficiency of the algorithm. Finally, following the recommendation of Refs.~\cite{Schubert2017,Sander1998}, we set \textit{minSamples} to 6, corresponding to twice the number of dimensions in the data space (here, three). 

\begin{figure}[h!]
    \centering
    \includegraphics[width=8cm]{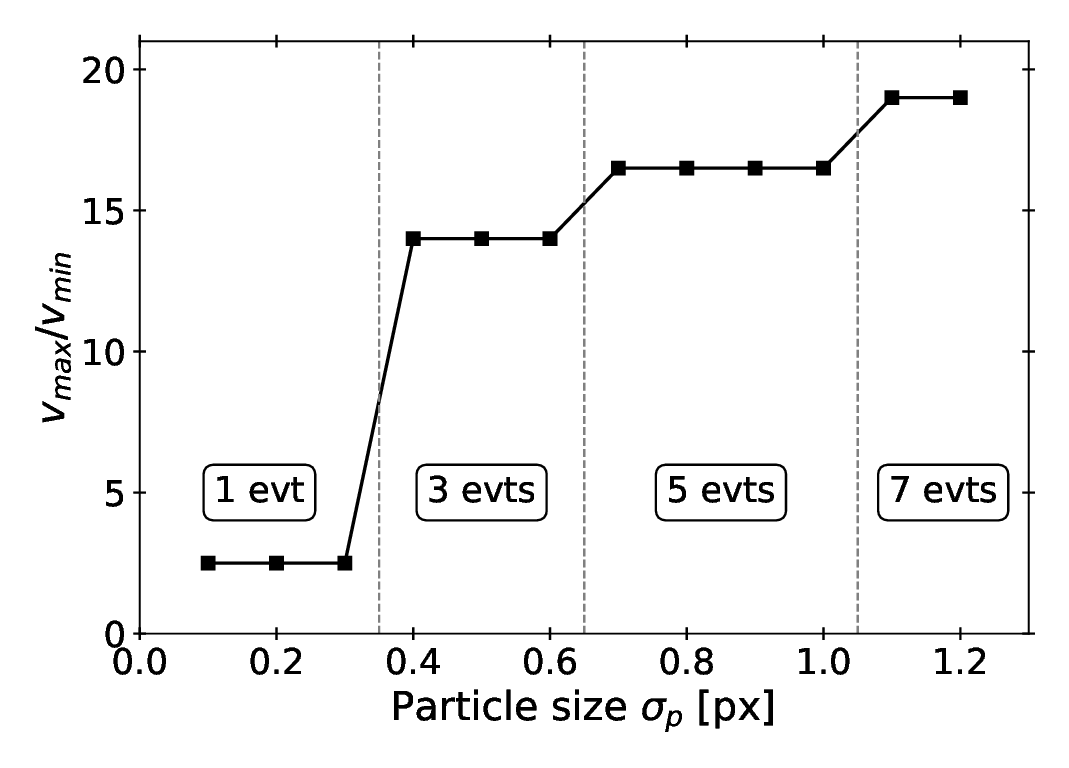}
    \caption{Velocity accessibility range for one set of DBSCAN parameters at $\ell = 10$~px as a function of the particle size $\sigma_p$. The boxes at the bottom of the figure indicate the corresponding event footprint widths associated with each $\sigma_p$ interval, delimited by vertical dashed lines.}
    \label{fig:VrangeVsSigma}
\end{figure}

To conclude this section, we further examine the width of the velocity range for which clustering succeeds. Figure~\ref{fig:VrangeVsSigma} shows the accessible velocity range, defined as the ratio $v_{\max}/v_{\min}$, at a fixed particle separation of $\ell = 10$~px, for various particle sizes $\sigma_p$. The corresponding event footprint widths for each $\sigma_p$ are indicated by the boxes at the bottom of the figure. When all other parameters are kept constant, the particle size influences the ratio $v_{\max}/v_{\min}$ in a discrete manner. This is due to the binary nature of the event data: changes in particle size are reflected only through the event footprint width, which can take only integer pixel values (here, 1, 3, 5, or 7). More importantly, Fig.~\ref{fig:VrangeVsSigma} shows that the accessible velocity range, already limited for large particles, decreases sharply as the particle size becomes smaller, reaching a ratio $v_{\max}/v_{\min} \approx 2.6$ in the smallest case. This ratio is extremely low, especially when compared to classical frame-based PIV, where values of $v_{\max}/v_{\min}$ up to 100 are typically achievable. In practice, for 2D2C flow visualization, small particles are desirable in order to achieve high seeding density, and therefore high spatial resolution, while avoiding saturation of the event-based sensor. This configuration represents then the worst case in terms of accessible velocity ratio. To overcome the limitations associated with small $\sigma_p$ and to exploit the intrinsic multi-time-scale nature of event-based acquisition, we have developed a new algorithm capable of achieving an arbitrarily large ratio $v_{\max}/v_{\min}$, even for closely spaced particles.

\section{Multiple time scale clustering}
\label{sec:clusterAlgo}

\subsection{The algorithm}
\label{sec:notreAlgo}

From the previous study, it appears that the range of velocities accessible with DBSCAN for a given set of parameters is too limited, particularly for small particles. To overcome this limitation, we developed a new algorithm that applies DBSCAN successively on the data for several values of $\alpha$, retaining only the clusters that satisfy specific acceptance criteria. 

\begin{figure}[h]
    \centering
    \includegraphics[width=7.5cm]{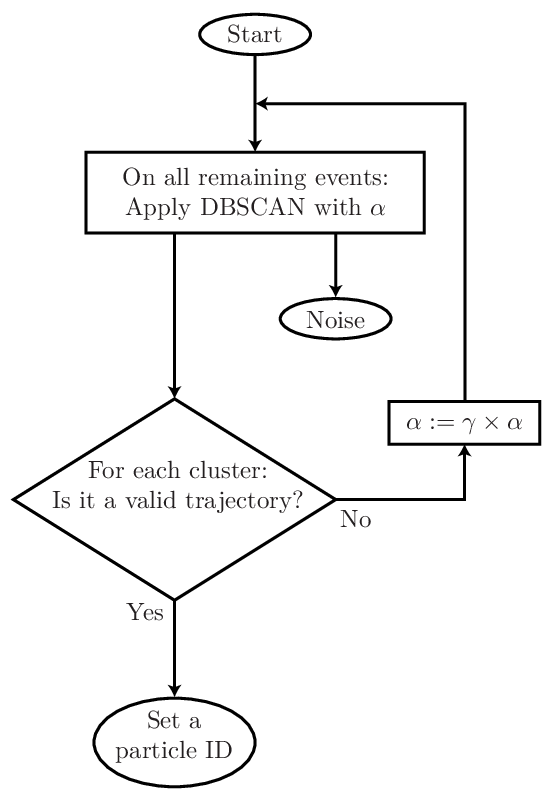}
    \caption{Schematic representation of the proposed algorithm. Here, $\alpha$ is the time-normalization factor in the Euclidean distance used by DBSCAN, see Eq.~(\ref{eq:metriqueAlpha}).}
    \label{fig:schemaAlgo}
\end{figure}

The algorithm operates as follows. We begin by applying DBSCAN to the dataset using the smallest possible value of $\alpha$, corresponding to the lowest expected particle velocity. Events discarded by DBSCAN are permanently considered as noise and removed from subsequent iterations. For each identified cluster, we then evaluate whether it satisfies a set of criteria defining a valid trajectory. If the criteria are met, the events of the cluster are assigned a cluster identification number. The events from rejected clusters are collected to form a new dataset, on which DBSCAN is applied again with $\alpha$ increased by a factor $\gamma$. The procedure continues iteratively until either a maximum value of $\alpha$ (corresponding to a maximum velocity) is reached, or no unclustered events remain. A schematic representation of this process is shown in Fig.~\ref{fig:schemaAlgo}. Based on the results from the previous section, particularly Fig.~\ref{fig:VrangeVsSigma} and its discussion, we recommend increasing $\alpha$ by a factor of $\gamma = 1.5$ between successive iterations. This factor can be adjusted: it may be increased when the event footprint width exceeds one pixel.

This algorithm therefore clusters each trajectory using the lowest acceptable value of $\alpha$ for it. This configuration also maximizes the number of events in the cluster, since a higher $\alpha$ increases the spatio-temporal distance between events and risks splitting the trajectory into several clusters.

The central question in this approach is which criteria should be used to retain only clusters corresponding to real particle trajectories. To address this, we recall the phase diagram in Fig.~\ref{fig:phaseDiagram}. When starting with the smallest $\alpha$, each cluster will correspond either to a valid trajectory (black region in Fig.~\ref{fig:phaseDiagram}) or to a combination of several trajectories (light blue region). Thus, the criterion must distinguish between a single trajectory and a bundle of overlapping ones. For each cluster, characterized by the triplet of arrays $(\mathbf{x}, \mathbf{y}, \mathbf{t})$ sorted in time, we propose to assess whether it represents a valid trajectory using the following procedure:

\begin{itemize}
\item[•] Clusters with fewer than \textit{minSamples} events cannot be considered as a valid trajectory.

\item[•] We compute the first-order temporal differences $\Delta \mathbf{t}$ between successive events. For the cluster to be considered a valid trajectory, the following condition must hold:
\begin{equation}
\text{std}(\Delta \mathbf{t}) / \text{mean}(\Delta \mathbf{t}) < \textit{diffTime} \, .
\end{equation}

\item[•] We perform a local quadratic fit over segments of \textit{windowsSize} events, with $50$~\% overlap. The final segment is computed using the last \textit{windowsSize} events. The fit $(\tilde{x}(t),\tilde{y}(t))$ is obtained via least-squares minimization. For each event $(x_n, y_n, t_n)$ in a segment, we compute the residual $res_n$, defined as the distance to the fitted trajectory:
\begin{equation}
res_n = \sqrt{(\tilde{x}(t_n)-x_n)^2 + (\tilde{y}(t_n)-y_n)^2} \,.
\end{equation}
The maximum residual among all events in all segments must remain below \textit{diffSpace} for the cluster to be considered a valid trajectory.
\end{itemize}

The first condition imposes a minimum number of events per cluster to avoid small, noisy clusters. It is worth recalling that DBSCAN can produce such small clusters containing fewer events than \textit{minSamples} when border points (i.e., events classified neither as core points nor as noise) are shared between core points that do not belong to the same cluster.

\begin{figure}[h]
\centering
\includegraphics[width=8cm]{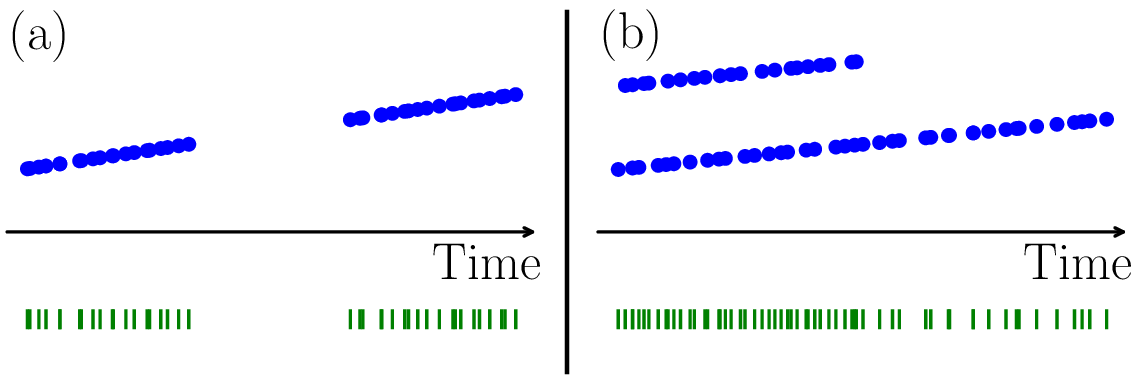}
\caption{Illustration of two situations where the temporal condition invalidates the cluster. In both panels, the cluster, composed of two trajectories, is plotted in blue along the vertical axis. At the bottom, green vertical lines indicate the event timestamps. On the left (a), the two trajectories are separated in time; on the right (b), they are spatially separated and one is shorter. In both cases, the temporal distribution is strongly non-uniform. For reference, the ratios $\text{std}(\Delta t)/\text{mean}(\Delta t)$ are approximately $2$ in (a) and $4$ in (b).}
\label{fig:critereTemps}
\end{figure}

The temporal condition enforces an approximately uniform temporal distribution of events, preventing clusters composed of particles appearing at different times but at the same location. It also helps to identify clusters formed by multiple overlapping trajectories, which typically exhibit a higher local event density due to the simultaneous presence of particles and the fine temporal discretization (see examples in Fig.~\ref{fig:critereTemps}). The procedure used to optimize the \textit{diffTime} parameter is described in Sec.~\ref{sec:parameterOpti}.

The last condition verifies that the cluster can be locally approximated by a quadratic fit. The number of events used for each fit, \textit{windowsSize}, increases linearly with the event footprint width of the trajectories. Here, we propose a value of $30$ for a three-event footprint width. A cluster is rejected if at least one event deviates from the fit by more than \textit{diffSpace} pixels. By construction, \textit{diffSpace} should be set slightly above half of the event footprint width.

\subsection{The two particles test for our algorithm}

\begin{figure}[h]
\centering
\includegraphics[width=8cm]{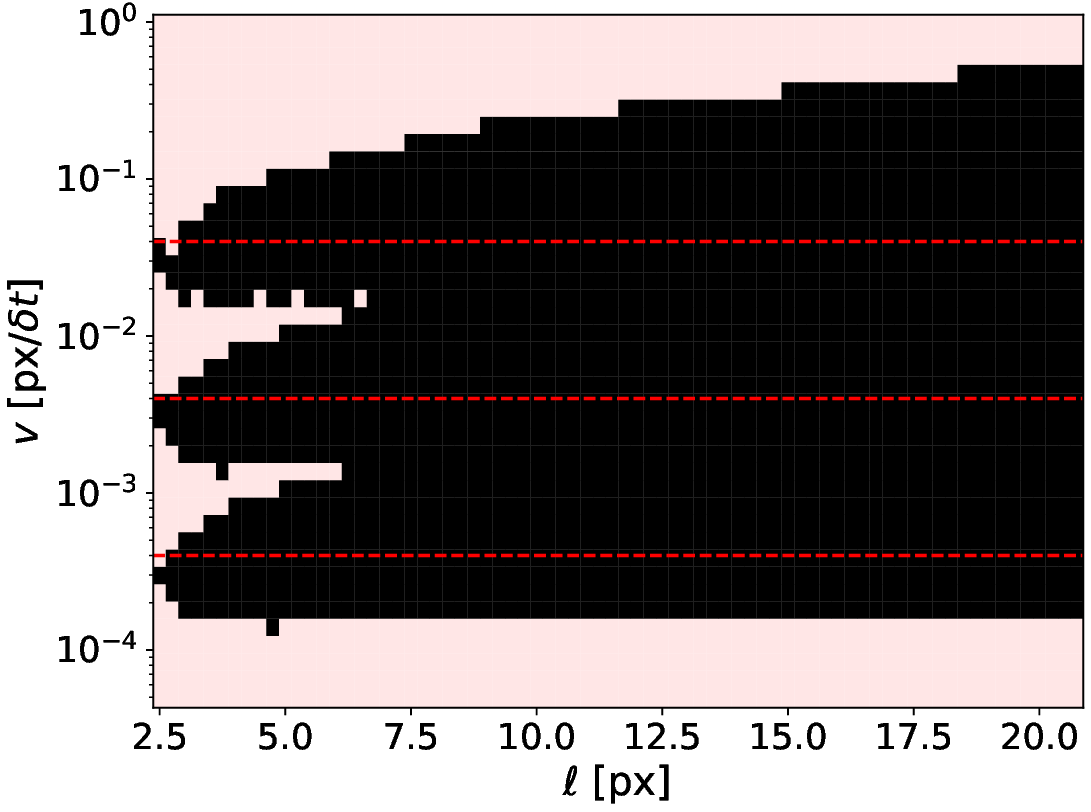}
\caption{Phase diagram of the two particle system clustered using the algorithm proposed in section~\ref{sec:notreAlgo}. Three values of time normalisation are tested, indicated by the red dashed lines: $\alpha = 4 \times 10^{-4}$, $4 \times 10^{-3}$, and $4 \times 10^{-2}$~px/$\delta t$.}
\label{fig:phaseDiagramVariousAlpha}
\end{figure}

Figure~\ref{fig:phaseDiagramVariousAlpha} shows the phase diagram obtained for the two particle test case using the clustering algorithm introduced above. In this implementation, we set $\textit{diffSpace} = 1.6$~px and $\textit{diffTime} = 1.2$, and selected three values of $\alpha$: $4 \times 10^{-4}$, $4 \times 10^{-3}$, and $4 \times 10^{-2}$~px/$\delta t$. All other simulation and DBSCAN parameters are identical to those listed in Tab.~\ref{tab:parameters2Particles}. As shown in the figure, the range of accessible velocities is greatly extended with the proposed algorithm, covering approximately three orders of magnitude in this configuration. The resulting domain of successful clustering corresponds to the union of the success regions obtained for each individual $\alpha$. More importantly, by increasing the range of $\alpha$ values tested by the algorithm, it becomes possible to achieve an arbitrarily large range of accessible velocities. Although the achievable range will ultimately be limited by hardware constraints (e.g., sensor saturation or timestamp resolution), the proposed method nevertheless represents a major improvement over the conventional approach and, in principle, even surpasses the performance of frame-based PIV, which is typically limited to a velocity range of $v_{\max}/v_{\min} < 100$. In other words, our algorithm inherently preserves, and effectively exploits, the multi–time-scale nature of event-based measurements.

\section{Multiple-particle tests}
\label{sec:multipleParticle}

\begin{figure}[h]
    \centering
    \includegraphics[width=8cm]{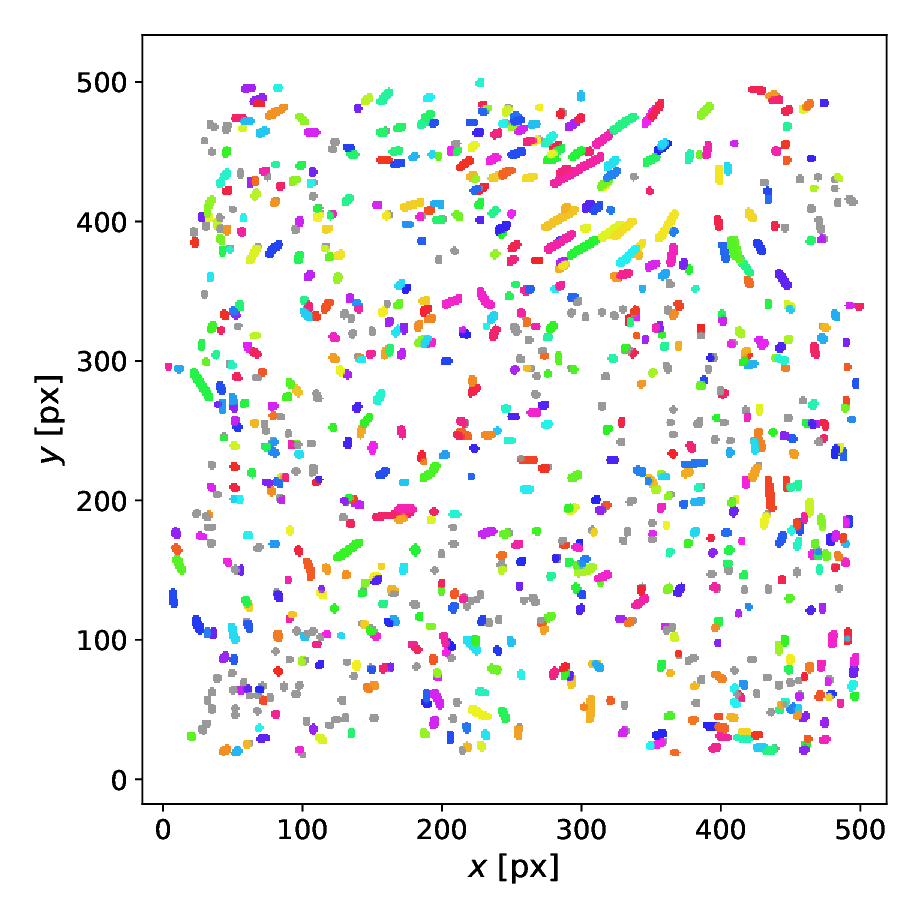}
    \caption{Illustration of the synthetic data in the $(x,y)$ plane, at low density used in this section clustered by our algorithm. Image created by accumulated 3000 time steps. As a reminder, events are processed in the space $(x,y,t)$. Events discarded by the algorithm are presented in gray. Due to the limited number of colors, separated cluster are sometimes represented with similar color.}
    \label{fig:IllustrationDataSynthetic}
\end{figure}

To fine-tune the parameters of the iterative DBSCAN algorithm proposed above and to evaluate its performance, we use a second set of synthetic data designed to be as realistic as possible. These data are generated from a set of Lagrangian trajectories obtained from a direct numerical simulation of homogeneous and isotropic turbulent flow. The simulation (already use in \cite{Valori2021}) was performed by Jean-Philippe Laval and correspond to a Reynolds number based on the Taylor microscale of $145$. The forcing is introduced through the Taylor-Green modes. To reproduce our 2D2C acquisition system, we select particles that take at least some values within a vertical plane of width $\Delta z/L = 0.5$~$\%$ in the third direction, where $L$ is the size of the simulated cubic domain. Particles within this plane are then randomly deselected to vary the particle density. The trajectories are linearly interpolated to increase the temporal resolution of the DNS by a factor of $20$, resulting in particle velocities always smaller than $0.01$~px/$\delta t$. Finally, the trajectories are projected onto a virtual sensor of $501 \times 501$~px${}^2$, and the corresponding set of events is generated as described in Section~\ref{sec:genereSyntheData}. In particular, no noise source is added during the generation process, and events are generated solely by particle motion. The values of the thresholds used for this dataset are $C_+ = C_- = 0.2$ and the signal-to-noise ratio is $I_{\max}/I_{\text{noise}} = 10$. Only positive-polarity events are processed. An illustration of the resulting data, at low density and clustered by the iterative DBSCAN algorithm presented above, is shown in Fig.~\ref{fig:IllustrationDataSynthetic}.

During the generation process, labeling events to indicate from which particle they originate is a challenging task. A first idea is to assume that an event is always generated by the particle closest to it. However, this approach is not satisfactory, since when particles are too close to each other, a single event may originate from several particles simultaneously. Moreover, negative-polarity events can be generated by particles that are no longer within the simulated plane. To label events in our synthetic data, we introduce a control radius, typically $5\sigma_p$, around each particle center. Pixels within this radius are considered to be ``influenced'' by the particle. When an event is generated by the pixel, the event is labeled as originating from all particles that have influenced this pixel since its last reset. A pixel is reset, and thus discards the list of particles that previously influenced it, when its light intensity drops to its minimum value $I_{\min}$.

\subsection{Evaluation criteria}

To properly evaluate the quality of the clustering produced by our algorithm, we introduce a new set of evaluation criteria. As a reminder, classical frame-based particle tracking velocimetry (PTV) evaluates performance on a frame-by-frame basis and distinguishes three relevant cases: a particle is present and detected (True Positive, TP), a particle is present but not detected (False Negative, FN), and no particle is present but a detection occurs (False Positive, FP). These cases lead to three number (\#TP, \#FN, and \#FP), from which two standard metrics are defined: the recall, defined as $\#\text{TP}/(\#\text{TP}+\#\text{FN})$, which answers the question ``How many particles are detected?'', and the precision, defined as $\#\text{TP}/(\#\text{TP}+\#\text{FP})$, which answers the question ``How many detections are related to true particle?''~\cite{Champagnat2014,LeBris2025}. The final evaluation is obtained by averaging recall and precision over several frames.

These classical evaluation criteria cannot be directly applied to our algorithm because event-based data do not provide discrete frames on which to base the analysis. Moreover, the method presented in this paper is a clustering-based approach whose objective is to recover entire particle trajectories rather than isolated particle detections. It is also worth noting that previous studies~\cite{Wang2020,AlSattam2024} on event-based particle-tracking evaluate their algorithms using velocity-based metrics. In our case, the velocity is not defined at this stage, and we prefer to focus on evaluating the event-to-trajectory assignment performance.

\begin{figure}[h]
    \centering
    \includegraphics[width=7.5cm]{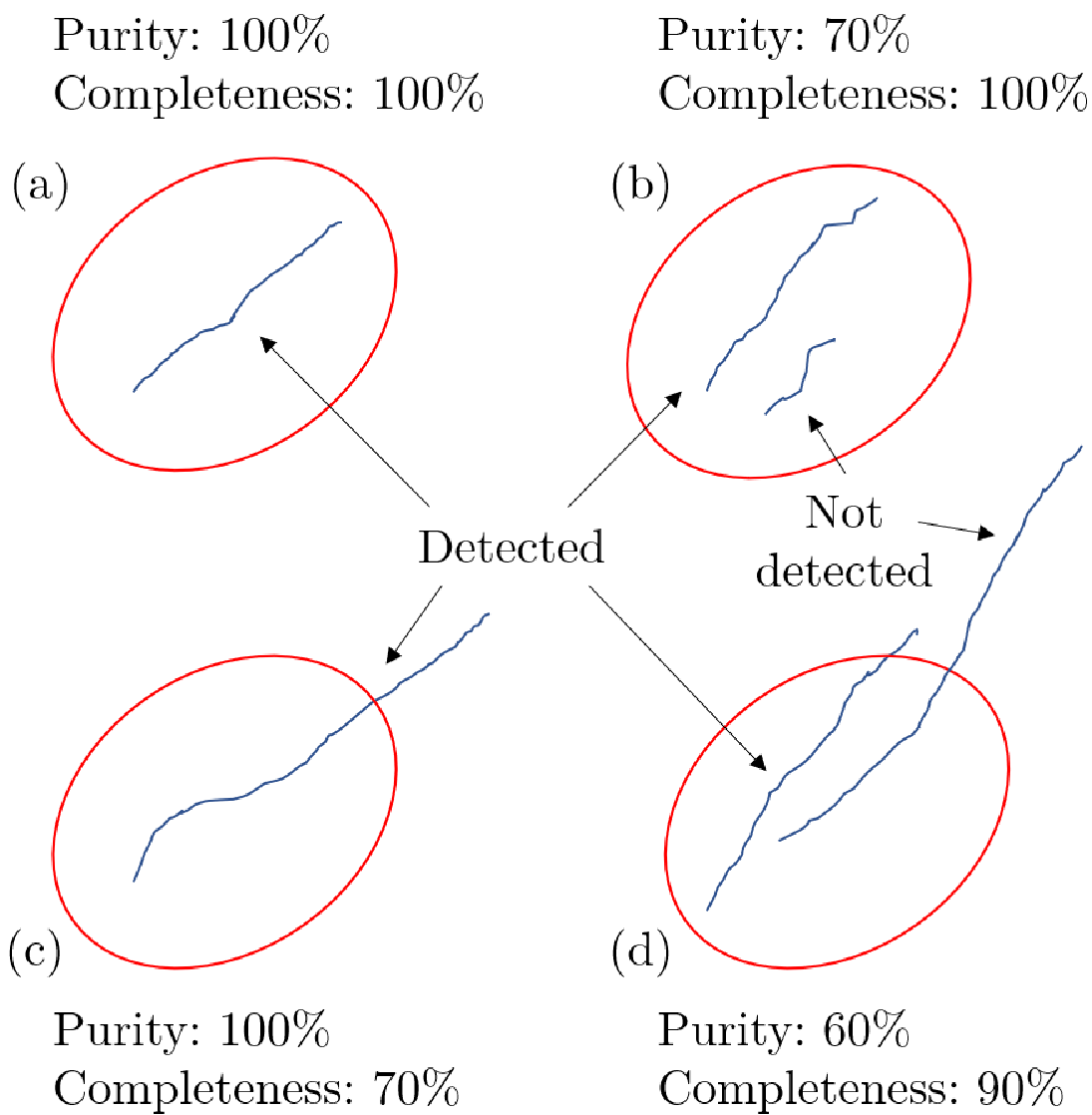}
    \caption{Schematic view of various clustering examples. Trajectories are shown in blue, and clusters are represented by red ellipses. Approximate values of purity and completeness are indicated for each cluster. For each cluster, one trajectory is detected. In clusters (b) and (d), another trajectory is not detected, as it is not the most represented trajectory in any cluster.}
    \label{fig:schemaEvaluation}
\end{figure}

Consequently, a new set of criteria is required to assess the quality of trajectory detection in event-based data. To this end, we propose focusing on three key questions: ``Is the trajectory detected?'', ``Does the cluster merge several trajectories?'', and ``Does the cluster cover the full trajectory?''. These questions rely on several important notions, which we define precisely below:
\begin{itemize}
    \item A trajectory is \textit{detected} if it is the most represented trajectory in at least one cluster. In other words, it is detected if there exists a cluster in which it contributes the largest number of events.
    \item The \textit{purity} of a cluster is defined as the fraction of its events originating from the most represented trajectory. Formally, the purity of a cluster $C$ is given by
    \begin{equation} \label{eq:purity}
        \text{Purity}(C) = \frac{|C \cap K|}{|C|}  \, ,
    \end{equation}
    where $|\cdot|$ denotes the number of events and $K$ is the reference trajectory that maximize the number of event within the intersection $C \cap K$ (i.e., the most represented trajectory in $C$). Purity is commonly used as an evaluation metric in the clustering community~\cite{Manning2009}.
    \item The \textit{completeness} of a cluster is defined as the fraction of events from the most represented trajectory that are contained in the cluster. Using the same notation as in Eq.~(\ref{eq:purity}), the completeness of a cluster $C$ is defined as
    \begin{equation}
        \text{Completeness}(C) = \frac{|C \cap K|}{|K|} \, .
    \end{equation}
\end{itemize}
Illustrative examples of these notions are provided in Fig.~\ref{fig:schemaEvaluation}. For events generated by multiple particles, each event is considered to belong to all corresponding trajectories. Consequently, such events are counted once for each trajectory when determining the most represented trajectory, its length, and its number of elements within the cluster.


Based on these notions, we define several evaluation criteria. The first is the \textit{recall}, defined as the fraction of detected trajectories. Trajectories with fewer than $6$~events are excluded, as they are too noisy to yield reliable Lagrangian velocity estimates.

Two other criteria are the mean purity and mean completeness, averaged over all clusters (hereafter referred to simply as \textit{purity} and \textit{completeness}). In addition, we compute the \textit{discarded rate}, defined as the fraction of events classified as noise by the algorithm. This last metric has the advantage of being computable experimentally, without requiring reference data, and can be used to tune algorithm parameters, as shown in the next section.

A good clustering algorithm for particle tracking should achieve a purity as high as possible, since impure clusters lead to noisy Lagrangian velocity estimates or may generate outliers. High completeness is also desirable, as it enables the detection of long trajectories, but it is less critical if the final objective is to compute an Eulerian velocity field. Finally, since no noise is added to our synthetic data, the discarded rate should vanish for a perfect algorithm.

\subsection{Parameter optimisation}
\label{sec:parameterOpti}

In this section, we propose to optimise the parameters used in the iterative DBSCAN algorithm introduced above. To this end, we use a synthetic dataset with a particle density of $0.004$~ppp and compute the evaluation metrics as functions of the parameters under investigation. For each configuration, two particle sizes are tested, $\sigma_p = 0.2$~px and $0.6$~px, corresponding to event footprints that are approximately 1 and 3 events wide, respectively.

\begin{figure}
    \centering
    \includegraphics[width=8cm]{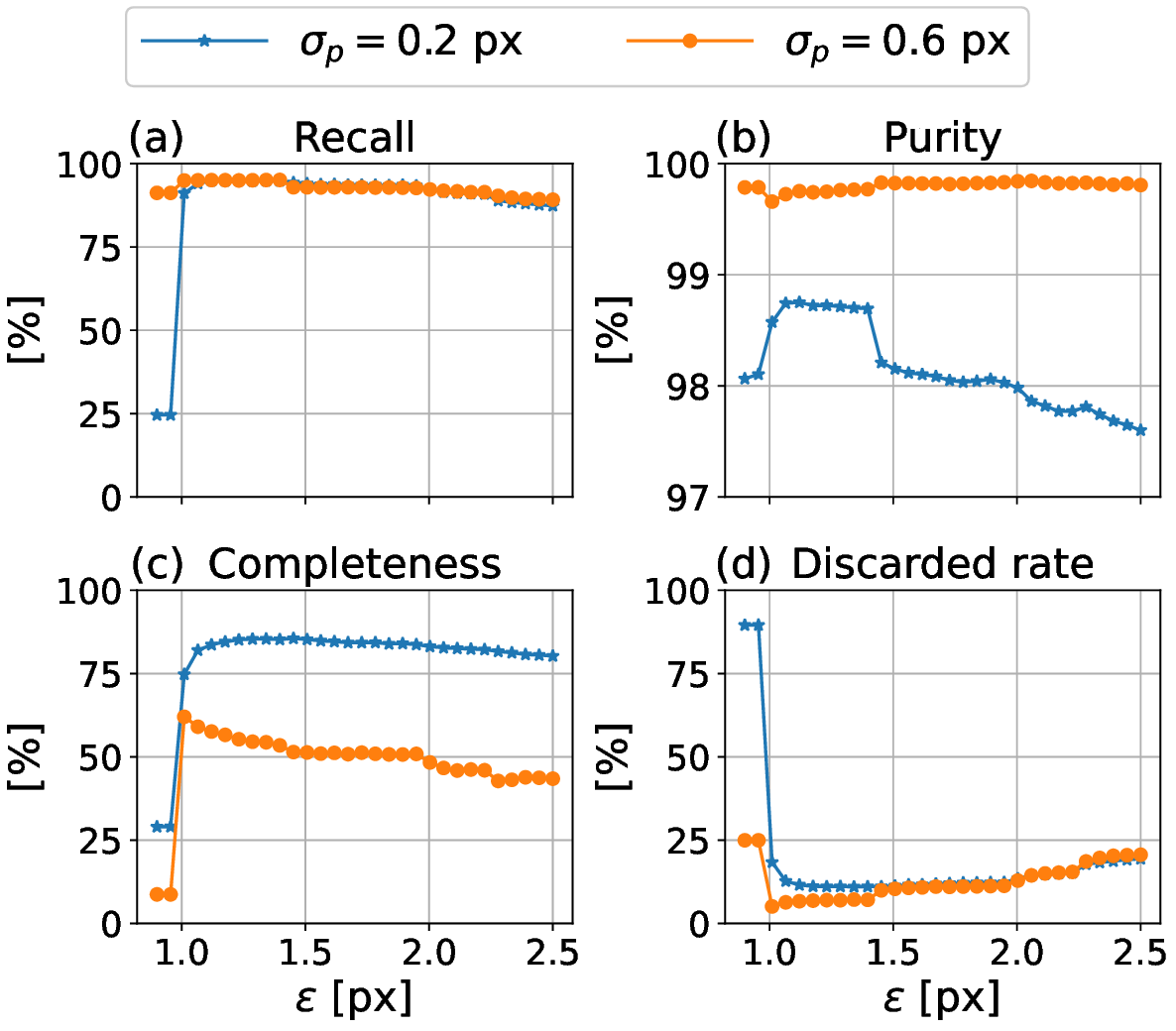}
    \caption{Evaluation criteria as a function of the neighborhood radius $\varepsilon$. Two particle sizes are tested: $\sigma_p = 0.2$~px (blue) and $0.6$~px (orange). The parameters \textit{diffSpace} and \textit{diffTime} are optimised in both cases and take the values $0.9$~px and $1.4$, respectively, for $\sigma_p = 0.2$~px, and $1.6$~px and $1.2$ for $\sigma_p = 0.6$~px.}
    \label{fig:variousEPS_syn}
\end{figure}

Figure~\ref{fig:variousEPS_syn} shows the evaluation criteria of the iterative DBSCAN algorithm as a function of the neighborhood radius $\varepsilon$. All other parameters are assumed to be optimised and depend on the particle size. We observe that for $\varepsilon < 1$~px, the algorithm performs very poorly, as expected, since events that do not lie within the same pixel cannot be clustered together. For $\varepsilon > 1$~px, the overall performance gradually degrades as $\varepsilon$ increases. Indeed, as noted in Section~\ref{sec:twoPartTest}, smaller values of $\varepsilon$ allow closer trajectories to be more effectively disentangled. The recall and the discarded rate exhibit clear, monotonic, and particle-size-independent trends with respect to $\varepsilon$. While the completeness for particles with $\sigma_p = 0.6$~px decreases with increasing $\varepsilon$, the completeness for $\sigma_p = 0.2$~px reaches a plateau in the range $\varepsilon \approx 1.3$--$1.8$~px. The purity, however, shows markedly different behavior depending on the particle size. For $\sigma_p = 0.6$~px, it remains approximately constant, with a slight increase around $\varepsilon = 1.5$~px followed by a small decrease. In contrast, for $\sigma_p = 0.2$~px, the purity exhibits a clear maximum close to $99$~\% in the range $\varepsilon = 1.1$--$1.4$~px. Since maintaining a high level of purity is crucial to preserve data quality and minimise the number of outliers, we select $\varepsilon = 1.4$~px for $\sigma_p = 0.2$~px particles (i.e., trajectories with a one-event-wide footprint), and $\varepsilon = 1.5$~px for $\sigma_p = 0.6$~px particles (i.e., trajectories with a three-event-wide footprint). These optimal values should be used in applications of the proposed algorithm and are expected to depend on the experimental conditions only through the event footprint width.

\begin{figure}
    \centering
    \includegraphics[width=8cm]{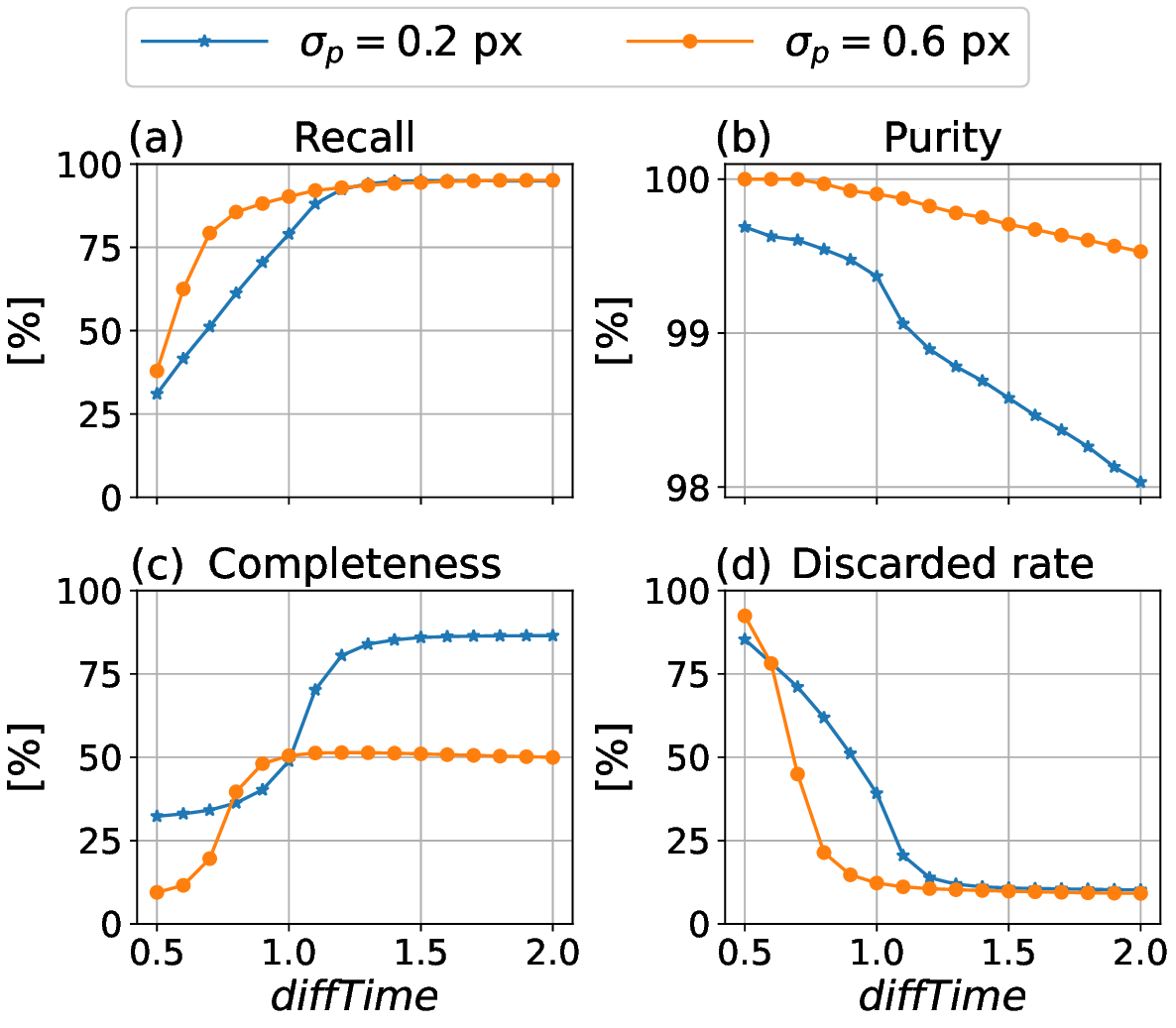}
    \caption{Evaluation criteria as a function of the \textit{diffTime} parameter. Two particle sizes are tested: $\sigma_p = 0.2$~px (blue) and $0.6$~px (orange). The parameters \textit{diffSpace} and $\varepsilon$ are optimised in both cases and take the values $0.9$~px and $1.4$~px for $\sigma_p = 0.2$~px, and $1.6$~px and $1.5$~px for $\sigma_p = 0.6$~px.}
    \label{fig:variousDiffTime}
\end{figure}

Similarly, Fig.~\ref{fig:variousDiffTime} shows the evaluation criteria as functions of the \textit{diffTime} parameter. The fine tuning of this parameter depends not only on the event footprint width, but also on the contrast thresholds $C_+$ and $C_-$, as well as on the signal-to-noise ratio $I_{\max}/I_{\min}$. Consequently, \textit{diffTime} must be tuned for each experimental configuration. The absolute values reported here are therefore less important than the methodology used to determine their optimal values. Figure~\ref{fig:variousDiffTime} shows that the purity decreases monotonically with \textit{diffTime}. In contrast, the recall, completeness, and discarded rate saturate above a threshold value of \textit{diffTime}, equal to approximately $1.4$ for $\sigma_p = 0.2$~px and $1.2$ for $\sigma_p = 0.6$~px. Below these thresholds, all three metrics depart rapidly from their saturation regime, indicating a strong degradation in algorithm performance. To minimise the number of outliers, we therefore seek to minimise \textit{diffTime} while keeping the other metrics within their saturation regime. In practice, when processing experimental data, \textit{diffTime} should be tuned by examining the discarded rate and selecting the smallest value corresponding to the onset of its plateau.

\subsection{Evaluation across various particle densities}

\begin{figure*}
    \centering
    \includegraphics[width=14cm]{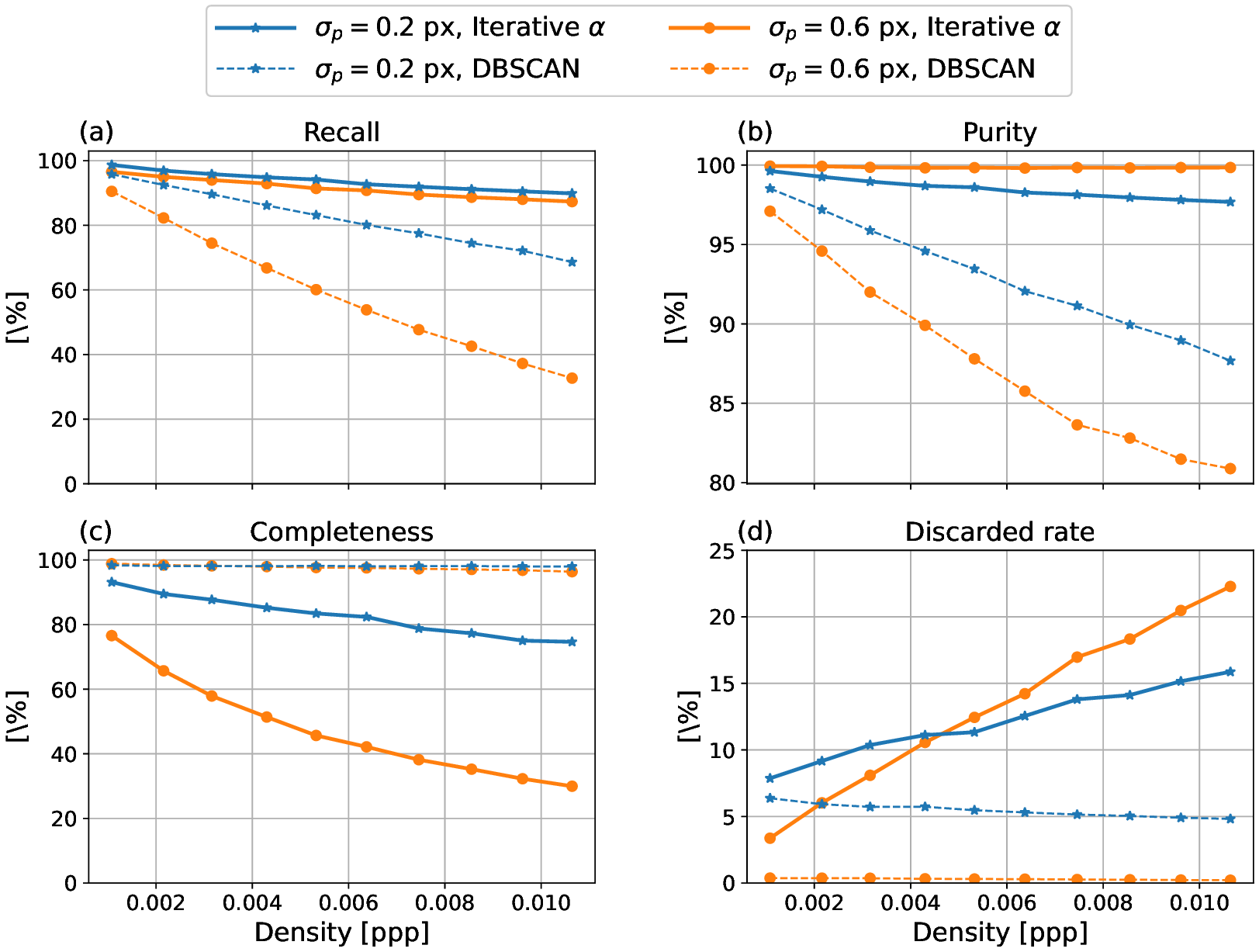}
    \caption{Evaluation criteria as functions of particle density. Two particle sizes are tested: $\sigma_p = 0.2$~px (blue) and $0.6$~px (orange). For each particle size, two algorithms are compared: the classical DBSCAN algorithm (dashed lines) and the iterative DBSCAN algorithm presented in Section~\ref{sec:clusterAlgo} (solid lines). All parameters are optimized for each configuration.}
    \label{fig:evalVSDensity}
\end{figure*}

Finally, we evaluate the performance of the algorithm over a range of particle densities from $0.001$ to $0.01$ particles per pixel (ppp). For reference, the particle density in the experimental data presented below is $0.004$~ppp. Figure~\ref{fig:evalVSDensity} shows the evaluation criteria as functions of particle density. Two particle sizes are considered, along with two clustering approaches: the classical DBSCAN algorithm and the iterative DBSCAN algorithm proposed in Section~\ref{sec:clusterAlgo}. All algorithm parameters are optimized for each particle size and each method.

Looking at Fig.~\ref{fig:evalVSDensity}, we observe that the proposed algorithm significantly improves both recall and purity. Although its performance degrades progressively as particle density increases, the recall still reaches $90$~\% at $0.01$~ppp, while the purity remains close to $100$~\% for $\sigma_p = 0.6$~px and above $97$~\% for $\sigma_p = 0.2$~px. Such performance is highly desirable for Lagrangian velocity estimation, as it enables reliable measurements for a large number of particles while minimizing outliers. The relatively low recall values obtained with DBSCAN can likely be explained by its limited accessible velocity range $v_{\max}/v_{\min}$, as discussed in Section~\ref{sec:twoPartTest}. Furthermore, based on Fig.~\ref{fig:evalVSDensity}, no sharp maximum particle density can be identified for the proposed algorithm; instead, its performance decreases gradually with increasing density. However, the iterative structure of the algorithm tends to fragment trajectories. This results in lower completeness than that obtained with DBSCAN, particularly for $\sigma_p = 0.6$~px. It also increases the discarded rate, as some trajectory segments may be too short to be considered valid trajectories. The tendency of the algorithm to fragment trajectories, especially when the event footprint width exceeds one event, constitutes an important limitation. In such cases, a post-processing step, similar to the ``Head and Tail Matching'' procedure proposed by~\cite{Hu2022}, may be required to reconnect fragmented trajectories.

\begin{figure}
    \centering
    \includegraphics[width=8cm]{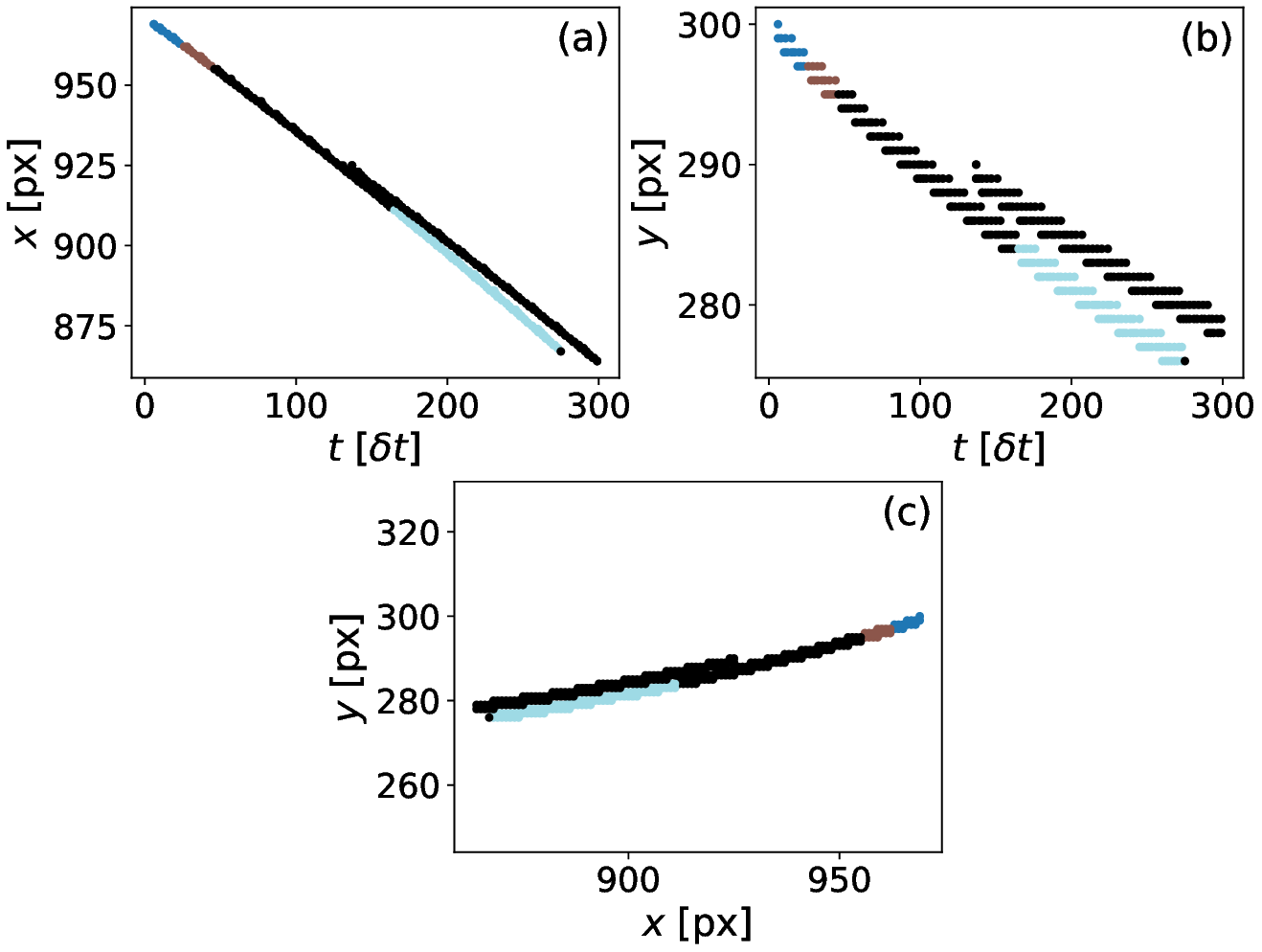}
    \caption{Spatio-temporal representation of two example trajectories clustered by the proposed algorithm. Panels (a) and (b) show $x$ and $y$ as functions of time, respectively, while panel (c) shows $x$ versus $y$. Discarded events are shown in black. The three clusters identified by the algorithm are shown in brown, blue, and light blue.}
    \label{fig:casEchec}
\end{figure}

To better understand why the proposed algorithm fragments trajectories, Fig.~\ref{fig:casEchec} presents an example of two trajectories that are very close to each other and that the algorithm fails to fully cluster. These trajectories are too close to be separated by DBSCAN using a single time normalization parameter $\alpha$. Rather than merging them into a single cluster---which would fail the cluster validation criteria---the iterative algorithm described in Section~\ref{sec:clusterAlgo} instead explores higher values of $\alpha$ and progressively segments the trajectories until extracting portions that satisfy the validation criteria. The remaining segments, corresponding to time intervals during which the particles are too close, are ultimately discarded. This strategy preserves cluster purity but increases the discarded rate and reduces completeness. 

Quantitative comparison with other state-of-the-art algorithms is challenging, as evaluation criteria are not yet standardized. Nevertheless, the review by \cite{AlSattam2024} compares several algorithms using synthetic data at higher particle densities than those considered here. Using evaluation criteria based on Lagrangian velocity vectors, they report that the best-performing algorithm is able to recover approximately half of the trajectories at a particle density of $0.03$~ppp. In contrast to our approach, their analysis does not address cluster purity or completeness. 

In the present case, since the experimental data analyzed below have a particle density of $0.004$~ppp, we consider the proposed algorithm to be sufficiently accurate to yield reliable results.

\section{Proof of concept on experimental data}
\label{sec:Expe}

Following the validation and parameter optimization performed on synthetic datasets, we now illustrate the applicability of the proposed clustering algorithm using experimental data acquired from a real turbulent flow. The experiments were carried out using a von Kármán apparatus. The device used here was built about two decades ago, and the turbulent flow it generates has been extensively characterized in previous works by members of our group. Detailed descriptions, schematics, and photographs can be found e.g. in Refs.~\cite{Cortet2010, Saint-Michel2013}. In brief, the von Kármán device consists of two counter-rotating impellers with curved blades of radius $R = 10$~cm, separated by a blade-to-blade distance of $14$~cm, and filled with water. In the configuration considered here, the impellers rotate in opposite directions in the scooping mode (the so-called ``ANTI'' configuration~\cite{Dubrulle2019}) at an identical frequency $f = 0.5$~Hz, generating a turbulent flow with a Reynolds number of $Re = 31000$. 

The flow is seeded with small polystyrene tracer particles (Spherotech) of diameter $5.9$~µm and density $1.05$~g/cm${}^3$, corresponding to a Stokes number based on the Kolmogorov scale of $St_\eta = 3 \times 10^{-4}$. To record the motion of these particles, we used a Prophesee EVK4-HD event-based camera powered by an IMX636 sensor, with a spatial resolution of $1280 \times 720$~pixels. Illumination was provided by a continuous-wave diode laser (Laser Tree LT-40W-AA) with an optical power of $5$~W, shaped into a laser sheet at the mid-plane of the von Kármán cell using a cylindrical lens. These low-cost diode lasers (less than 200~\euro{}) are sufficient for flow visualization, thanks to the high sensitivity of event-based cameras. Table~\ref{tab:parametersExpe} summarizes the experimental parameters used in this study, as well as the camera bias settings and the parameters used in the processing algorithm. The camera bias values are normalized by the manufacturer. The thresholds \textit{diff\_on} and \textit{diff\_off} correspond to $C_+$ and $C_-$. However, no interpretation in terms of relative light intensity is provided by the manufacturer. Regarding the algorithm settings, we set \textit{minSamples} to $6$ and convert $\varepsilon$ and \textit{diffSpace} into millimeters using the mean conversion factor \textit{px\_to\_mm} obtained from calibration and distortion correction.

\begin{table*}[t]
  \centering
  \begin{tabularx}{\textwidth}
  {>{\centering\arraybackslash}X
   >{\centering\arraybackslash}X|
   >{\centering\arraybackslash}X
   >{\centering\arraybackslash}X|
   >{\centering\arraybackslash}X
   >{\centering\arraybackslash}X}
  \toprule
  Experimental parameters & Value & Camera bias & Value  & Clustering setting & Value \\
  \midrule
  $f$ & 0.5 Hz & \textit{diff\_on} & 190 & $\gamma$ & 1.5 \\
  $Re$ & 31\,000 & \textit{diff\_off} & 84 & $\varepsilon$ & 0.125 mm \\
  Event rate & 3 Mevt/s & \textit{bias\_hpf} & 43 &  \textit{diffSpace} & 0.18 mm \\
  Particle density & 0.004 ppp & \textit{bias\_fo} & 0 & \textit{diffTime} & 1.1 \\
  $\eta$ & 90~µm & \textit{bias\_refr} & 0 & \textit{windowsSize} & 30   \\
  $\tau_\eta$ & 8~ms & -- & -- & \textit{minSamples} & 6  \\
  $St_\eta$ & $3 \times 10^{-4}$ & -- & -- & \textit{px\_to\_mm} & 12.2 px/mm \\
  \bottomrule
  \end{tabularx}
  \caption{Experimental parameters, camera bias and algorithm setting. Here $f$ is the frequency of the impeller, $Re = 2 \pi R^2 f/\nu$ is the Reynolds number. The mean event rate (with $1\,\text{Mevt/s} = 10^6\,\text{events/s}$) includes both positive and negative events, although only positive events are processed. Kolmogorov length and time scales are denoted $\eta$ and $\tau_{\eta}$ and the Stokes number based on $\tau_\eta$ is $St_\eta$. Details about the camera bias can be found on the manufacturer website. Discussion about the interpretation of the algorithm settings can be found in Sections~\ref{sec:clusterDBSCAN} and~\ref{sec:clusterAlgo}. \textit{px\_to\_mm} is the mean pixel-to-millimeter conversion ratio.}
  \label{tab:parametersExpe}
\end{table*}

The experimental data are calibrated using a static calibration target printed from an online generator (white dots on a black background). Such a static target should not normally be detected by an event-based camera; however, by using an oscillating light source (here, the room’s neon lighting) and setting the event-based thresholds to low values, we are able to detect events generated by the static target. From these data, we divide the dataset into time windows of $100$~ms and reconstruct images by assigning each pixel a value of 1 if at least one event is recorded at that location, and 0 otherwise. We then apply the open-source Python library OpenCV to the image obtained by time-averaging these reconstructed binary images. OpenCV performs geometric camera calibration, estimating both radial and tangential distortion coefficients. Finally, the coordinates of all events are converted from pixel units to physical units (mm) before further processing.

\begin{figure}[h]
    \centering
    \includegraphics[width=8cm]{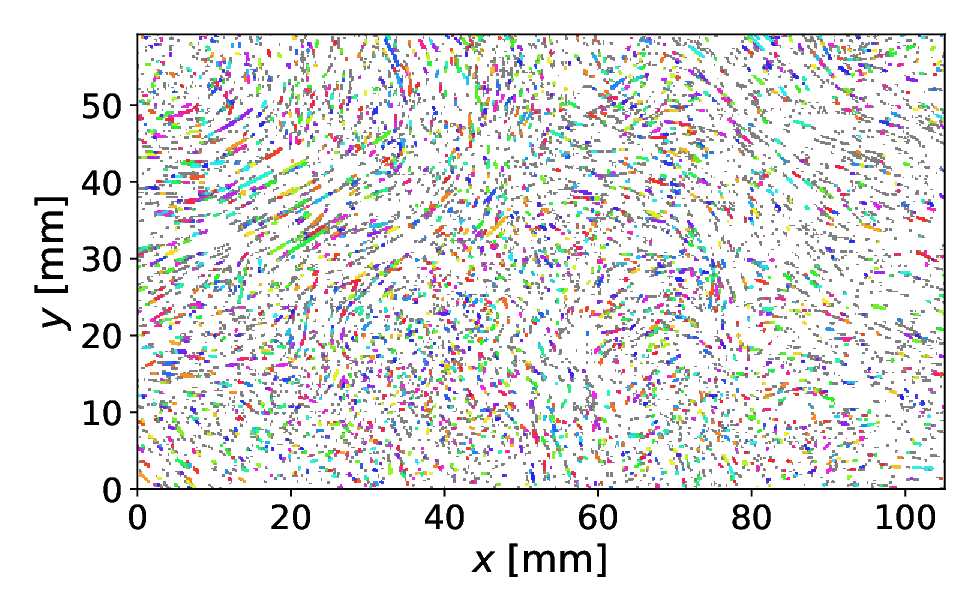}
    \caption{Illustration of event clustering on experimental data over the full measurement field. Approximately 80~000 positive events are shown, corresponding to 50~ms of acquisition. Events classified as noise are displayed in gray.}
    \label{fig:illustationExp}
\end{figure}

Figure~\ref{fig:illustationExp} illustrates the results produced by the proposed clustering algorithm when applied to the experimental data. Approximately $24$~\% of the positive events recorded by the event-based camera are discarded by the algorithm. This relatively high discarded rate arises from both background noise in the experimental data and a non-negligible number of valid trajectories that are partially or entirely discarded by the algorithm. Indeed, effective experimental procedures to further reduce background noise remain unclear. In the present experiment, we attempted to minimize noise by tuning the event-based thresholds to their maximum values and adjusting the high-pass filtering. Despite the use of water filtered with a $3$~µm filter, preliminary tests performed before adding particles revealed that the camera still recorded an event rate of $0.1$~Mevt/s. Moreover, although the proposed algorithm clearly discards or fragments some valid trajectories, we did not observe any clusters that appear to merge multiple trajectories.

\begin{figure}[h]
    \centering
    \includegraphics[width=8cm]{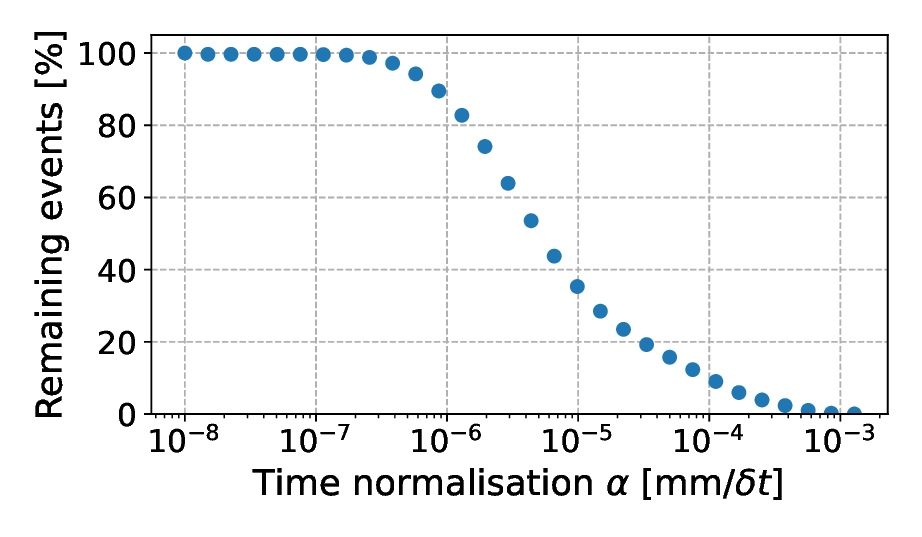}
    \caption{Remaining events to be clustered by the iterative algorithm presented in section~\ref{sec:clusterAlgo} versus $\alpha$. The algorithm stops when this number drops to zero.}
    \label{fig:avancementExp}
\end{figure}

Figure~\ref{fig:avancementExp} shows the evolution of the percentage of events remaining in the main loop of the algorithm as a function of the time normalization parameter $\alpha$. The initial value is set to $\alpha_0 = 10^{-8}$~mm/$\delta t$, and $\alpha$ is increased by a factor $\gamma = 1.5$ between two successive iterations of the main loop. The initial value $\alpha_0$ should be sufficiently small to correctly detect slow particles. However, choosing $\alpha_0$ excessively small increases the computational cost. As shown in Fig.~\ref{fig:avancementExp}, approximately three decades of $\alpha$ values are required to fully process the data, indicating a wide range of particle velocities in the experimental flow. The algorithm terminates when all events have either been assigned to a cluster or discarded.

\begin{figure}[h]
    \centering
    \includegraphics[width=8cm]{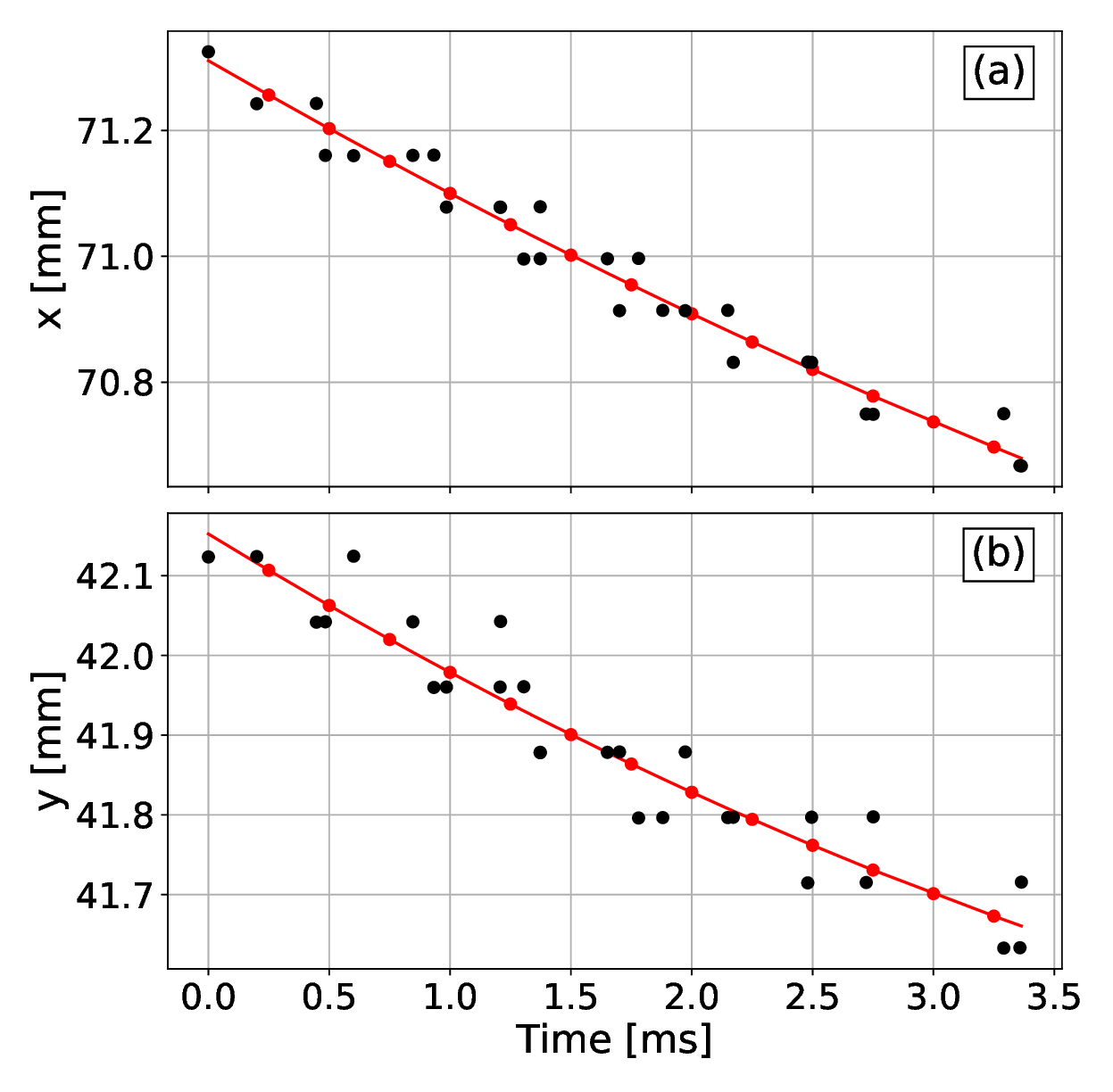}
    \caption{Illustration of a quadratic fit applied to a single particle trajectory. Panel~(a) shows the fit along the horizontal $x$ direction, and panel~(b) along the vertical $y$ direction. The fit is performed using a least-squares method. Red dots indicate the times of the global temporal grid at which the particle position and velocity are evaluated.}
    \label{fig:fit_quadra}
\end{figure} 

\begin{figure}[h]
    \centering
    \includegraphics[width=8cm]{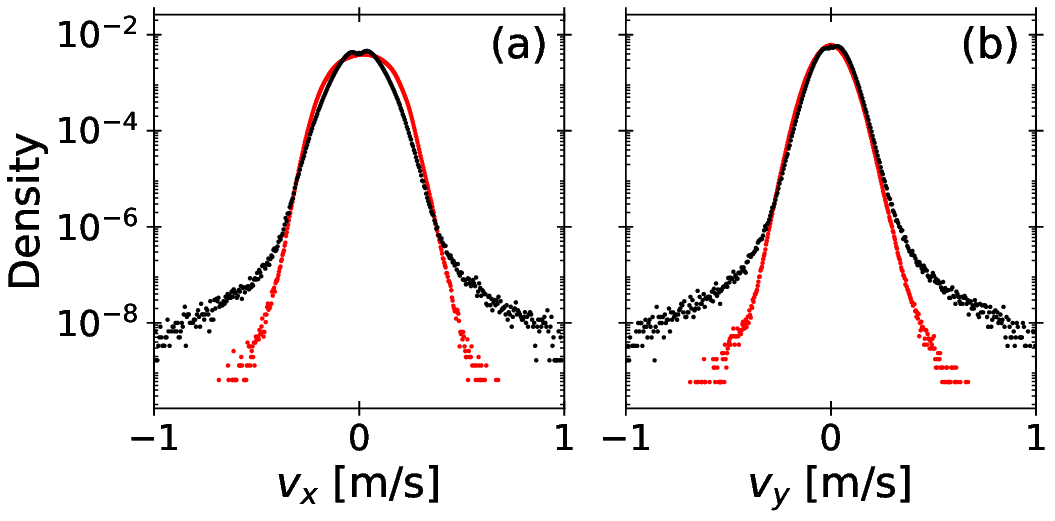}
    \caption{Probability density functions of the velocity components $v_x$ and $v_y$ obtained from our event-based recording system (black) and from a previous study using a 3D frame-based acquisition system~\cite{Debue2021}, processed with the Shake-The-Box algorithm~\cite{Schanz2016} (red). Because the frame-based acquisition used a smaller observation window than the present study, the PDFs are not expected to be strictly identical.}
    \label{fig:distribVitesse}
\end{figure}

Finally, once clustering is completed, we can measure the Lagrangian velocity of each cluster. Since our measurements correspond to a three-dimensional turbulent flow observed within a two-dimensional vertical plane, most particles simply cross the laser sheet and leave a quasi-linear footprint. We therefore compute the Lagrangian velocity of each cluster by performing a quadratic fit of the particle position as a function of time, using all events belonging to the cluster. The choice of a quadratic fit for trajectory reconstruction is motivated by the relatively simple shape of the measured trajectories, which are almost linear in the present experiments. A quadratic model is therefore sufficient to capture the particle motion while allowing for small deviations from linearity. This global fit allows the particle position to be estimated with sub-pixel accuracy, thereby avoiding peak-locking effects. Using the coefficients of the quadratic fit, the particle position and velocity are then evaluated on a common global time grid, enabling direct comparison between particles. This global time grid starts at zero, and its time step, denoted \textit{globalTimeStep}, is set here to $250$~µs. Adjusted positions and velocities are computed only for global time steps that lie strictly within the temporal extent of the cluster. To improve data quality, we further discard trajectories with an average temporal spacing larger than $4$~ms per event and trajectories whose velocity magnitude $v = \sqrt{v_x^2+v_y^2}$ exceeds $1000$~mm/s.

An illustration of such a quadratic fit and the resulting adjusted positions is shown in Fig.~\ref{fig:fit_quadra}. A non-negligible dispersion of the events around the fitted trajectory can be observed, which arises from the absence of sub-pixel positional information. This limitation may lead to inaccuracies in velocity estimation for short trajectories and restricts reliable measurements of low vertical or horizontal velocities. Such effects are particularly pronounced for nearly vertical or nearly horizontal trajectories, where particle displacements span only a small number of pixels in at least one direction. Moreover, since positive events are generated only at the particle boundary on the leading edge (i.e., in the direction of motion), the fitted position does not correspond exactly to the particle center. Further developments are therefore required to rigorously reconstruct particle trajectories from the positive-polarity events they generate.

In Fig.~\ref{fig:distribVitesse}, we show the probability density functions (PDFs) of the measured Lagrangian velocity components $v_x$ and $v_y$. For comparison, we also report data obtained in the same experimental configuration in a previous study conducted by our group using a three-dimensional frame-based acquisition system~\cite{Debue2021}. These reference data were processed using the Shake-The-Box algorithm~\cite{Schanz2016} within an observation domain of $5$~cm $\times$ $3.5$~cm $\times$ $0.67$~cm. Only particles located within the central millimeter along the third spatial direction are shown here. Because the observation domain of the present study is different and approximately twice as large as that of the frame-based measurements, the resulting PDFs are not expected to be strictly identical.

As shown in Fig.~\ref{fig:distribVitesse}, the central regions of the velocity PDFs obtained from the event-based measurements are in overall good agreement with the reference frame-based data, indicating that the proposed algorithm captures the dominant Lagrangian velocity statistics of the flow. Small discrepancies are nevertheless observed in the central region of the horizontal velocity component $v_x$. These discrepancies arise from the sensitivity of the velocity statistics to the observation domain and from the absence of sub-pixel positional information, which limits accurate estimation of low velocities when particle displacements span only one or two pixels along a given direction, as is the case for nearly vertical trajectories. In addition, the smooth and continuous shape of the PDFs confirms the absence of peak-locking effects in the reconstructed velocities. Differences become more pronounced for larger and less probable velocities, where the PDFs obtained from the event-based processing exhibit broader tails. This over-representation of large velocities is mainly attributed to the fragmentation of some trajectories by the clustering algorithm. When a trajectory is split into several shorter clusters, the velocity estimation is performed over a reduced temporal interval, making it more sensitive to local position uncertainties and potentially leading to an overestimation of the instantaneous velocity. It is furthermore worth recalling that event-based particle tracking remains a relatively recent research field, and its performance is not yet expected to match that of state-of-the-art frame-based algorithms, which benefit from several decades of development.

\begin{figure}
    \centering
    \includegraphics[width=8cm]{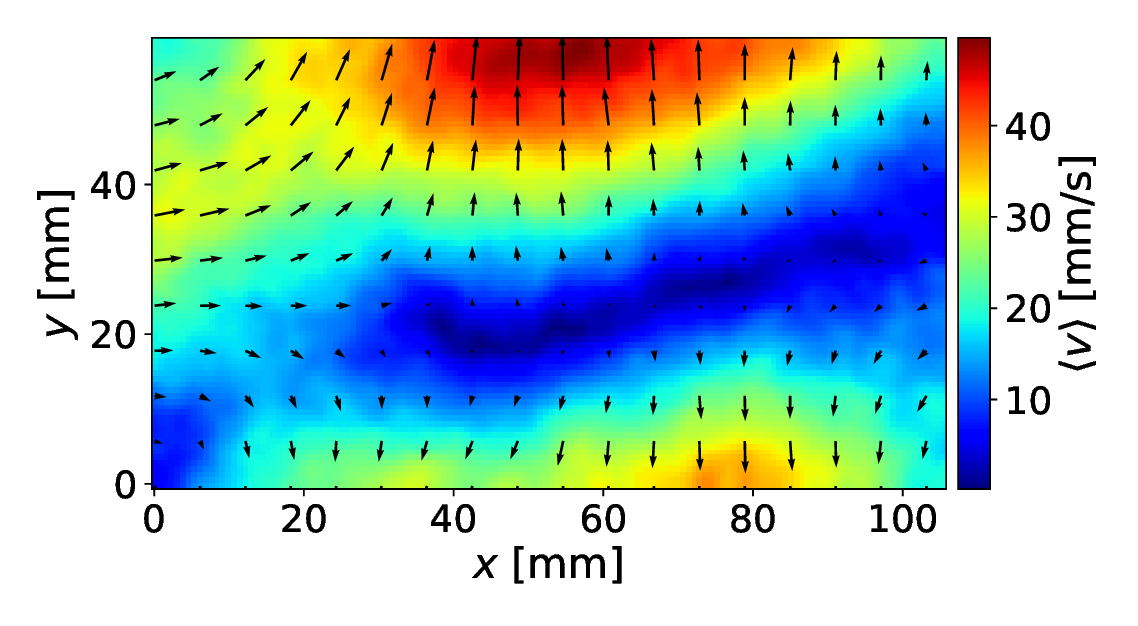}
    \caption{Eulerian mean velocity field computed from Lagrangian particle trajectories. The color map shows the magnitude of the mean velocity, while arrows indicate the mean flow direction.}
    \label{fig:meanFields}
\end{figure}

Finally, we compute the Eulerian mean flow from the Lagrangian trajectories. The particle velocities within each interrogation window are averaged to obtain the Eulerian mean flow, and a Gaussian filter is then applied for smoothing. The resulting fields are shown in Fig.~\ref{fig:meanFields}. The mean flow is in qualitative agreement with previously reported measurements of the same von K\'arm\'an configuration~\cite{Dubrulle2019}. In particular, we observe the four-cell structure of the mean flow, delimited by $x \approx 60$~mm and $y \approx 30$~mm, which is characteristic of a von K\'arm\'an device.

\section{Conclusion}

In this paper, we investigated the use of the DBSCAN algorithm for particle tracking with event-based data. Using a simple two-particle system, we showed that the classical DBSCAN algorithm is too limited in terms of accessible velocity range to be used in practice, especially when particle event footprints are one event wide. To overcome this limitation, we proposed a new clustering-based particle-tracking algorithm relying on an iterative application of DBSCAN. A key feature of this algorithm is that it does not assume any characteristic time scale or velocity scale, thereby preserving the multi-timescale nature of event-based data and enabling the detection of particles over an arbitrarily wide velocity range. The proposed algorithm uses a set of validation criteria to determine which clusters are retained. These criteria ensure both the temporal uniformity of the event distribution along a trajectory and its spatial consistency. The performance of the algorithm was evaluated using synthetic datasets. Particular attention was paid to the definition of evaluation metrics for trajectory detection adapted to event-based data. These metrics are based on three complementary questions: ``Is the trajectory detected?'', ``Does the cluster merge several trajectories?'', and ``Does the cluster cover the full trajectory?''. These questions lead to the definition of three evaluation metrics ---recall, purity and completeness--- to which we added the discarded rate. The algorithm parameters were optimized with respect to these metrics, with particular emphasis on maximizing cluster purity, even at the cost of increased trajectory fragmentation. The performance of the algorithm was evaluated for particle densities ranging from $0.001$ to $0.01$~ppp. Finally, the algorithm was applied to experimental data recorded in a turbulent flow generated by a von Kármán device. This application demonstrates that the proposed method can be successfully applied to real experimental data. Although its performance does not yet reach that of state-of-the-art frame-based algorithms, the results confirm the relevance and potential of clustering-based approaches for event-based particle tracking.

Several aspects of the algorithm could be improved in future work. First, negative-polarity events should be more effectively exploited. In the experiments, they account for approximately half of the recorded data but are not currently processed. This second channel contains information similar to that of positive-polarity events, albeit slightly shifted in space. For example, Hu \textit{et al.}~\cite{Hu2022} proposed to consider an event as a core point only if at least 90~\% of the events in its neighborhood share the same polarity. A second limitation concerns the lack of directionality in the clustering process. When DBSCAN searches within the neighborhood of an event, all neighboring events are treated equally, despite the fact that the global shape of a trajectory often exhibits a clear preferential direction. This can lead to failure cases such as those illustrated in Fig.~\ref{fig:casEchec} and limits the achievable particle density. Moreover, the absence of directionality currently prevents straightforward extension to three-dimensional applications, where trajectories are expected to cross in space and where events may be generated by multiple particles.

\section*{Acknowledgments}

This work was supported by a French government grant managed by the Agence Nationale de la Recherche France 2030 under Reference No. ANR-24-RRII-0001. The authors would like to thank Cécile Wiertel-Gasquet, Vincent Padilla, Frédéric Coneggo, and the entire technical team of the SPEC laboratory for their experimental support. Special thanks to Jean-Philippe Laval from LMFL, for providing us with the Direct Numerical Simulation results used in section~\ref{sec:multipleParticle}. The authors would also like to thank Laurent Soulier for his proofreading and comments.

\section*{Code availability}

The code developed in this study and the datasets used for the analysis are available from the corresponding author upon reasonable request.


\begin{thebibliography}{99}

\bibitem{Cortet2010} Cortet P-P, Chiffaudel A, Daviaud F, and Dubrulle B (2010), Experimental Evidence of a Phase Transition in a Closed Turbulent Flow. Phys. Rev. Let. 105, 214501

\bibitem{Saint-Michel2013} Saint-Michel B, Dubrulle B, Marie L, Ravelet F, and Daviaud F (2013) Forcing-type-dependent stability of steady states in a turbulent swirling flow. Phys. Rev. Let. 111,  234502

\bibitem{Dubrulle2019} Dubrulle B (2019), Beyond Kolmogorov cascades. J. Fluid. Mech. (867)

\bibitem{Obligado2013} Obligado M, Puy M, and Bourgoin M (2013). Bi-stability of a pendular disk in laminar and turbulent flows. J. Fluid Mech. 728, 1-11 

\bibitem{Gayout2021} Gayout A, Bourgoin M, and Plihon N (2021). Rare Event-Triggered Transitions in Aerodynamic Bifurcation. Phys. Rev. Lett. 126, 104501

\bibitem{Lanchon2023} Lanchon N, Mora DO, Monsalve E, and  Cortet PP (2023) Internal wave turbulence in a stratified fluid with and without eigenmodes of the experimental domain, Phys. Rev. Fluids 8:054802

\bibitem{Lanchon2025} Lanchon N, Boury S,  and Cortet PP (2025) Laboratory observation of internal gravity wave turbulence in a three-dimensional large-scale facility, Phys. Rev. Fluids 10:084804

\bibitem{Gallego2020} Gallego G, Delbruck T, Orchard G, Bartolozzi C, Taba B, Censi A, Leutenegger S, Davison A, Conradt J, Daniilidis K, Scaramuzza D (2020) Event-based Vision: A Survey. IEEE Trans. Pattern Anal. Mach. Intell. 44(1):154-180

\bibitem{McReynolds2022} McReynolds B, Graca R, Delbruck T (2022) Experimental methods to predict dynamic vision sensor event camera performance. Opt. Eng. 61(7):074103

\bibitem{Graca2021} Graca R, Delbruck T (2021) Unraveling the paradox of intensity-dependent DVS pixel noise. Workshop: IISW, 2021

\bibitem{Willert2022} Willert C, Klinner J (2022) Event-based imaging velocimetry: An assessment of event-based cameras for the measurement of fluid flows. Exp. Fluids 63:101

\bibitem{Lichtsteiner2008} Lichtsteiner P, Posch C, Delbruck T (2008) A 128$\times$128 120 dB 15 µs Latency Asynchronous Temporal Contrast Vision Sensor. IEEE J. Solid-State Circuits. 43(2):566-576 

\bibitem{Posch2011} Posch C and Matolin D (2011) Sensitivity and uniformity of a 0.18µm CMOS temporal contrast pixel array,  IEEE International Symposium of Circuits and Systems (ISCAS), pp. 1572-1575

\bibitem{Howell2020} Howell J, Hammarton TC, Altmann Y, Melanie Jimenez M (2020) High-speed particle detection and tracking in microfluidic devices using event-based sensing, Lab Chip 20 (16)

\bibitem{Willert2023} Willert C (2023) Event-based imaging velocimetry using pulsed illumination. Exp. Fluids 64:98

\bibitem{Willert2025} Willert C, Klinner J (2025) Dynamic Wall Shear Stress Measurement using Event-based 3D Particle Tracking. Exp. Fluids 66:32

\bibitem{Franceschelli2025} Franceschelli L, Willert CE, Raiola M, Discetti S (2025) An assessment of event-based imaging velocimetry for efficient estimation of low-dimensional coordinates in turbulent flows, Exp. Therm. Fluid Sci 164, 111425

\bibitem{Drazen2011} Drazen D, Lichtsteiner P, Häfliger P, Delbruck T, Jensen A (2011) Toward real-time particle tracking using an event-based dynamic vision sensor. Exp Fluids 51:1465–1469

\bibitem{Borer2017} Borer D,  Delbruck T, Rösgen T (2017) Three‑dimensional particle tracking velocimetry using dynamic vision sensors. Exp. Fluids 58, 165

\bibitem{Rusch2023} Rusch A and Rösgen T (2023) TrackAER: real-time event-based quantitative flow visualization. Exp. Fluids 64, 136

\bibitem{AlSattam2024} AlSattam O, Mongin M, Grose M, Gunasekaran S, Hirakawa K (2024) KF‑PEV: a causal Kalman filter‑based particle event velocimetry. Exp. Fluids 65:141

\bibitem{Hu2022} Hu S, Kim Y, Lim H, Lee AJ, Myung H (2022) eCDT: Event Clustering for Simultaneous Feature Detection and Tracking.  IEEE/RSJ International Conference on Intelligent Robots and Systems (IROS):3808-3815

\bibitem{Bera2026} Bera S, Bao D, Schweitzer F, Vienney C, Lanchon N, Cheminet A, Dubrulle B, Bourgoin M, Plihon N (2026) Experimental Characterization of Quadrotor Flow Topology Near the Ground Using Event-Based Stereo-PIV, AIAA SciTech

\bibitem{Sachs2026} Sachs S, Jung S, Kahl M, Keuper M, Willert C, Cierpka C (2026) STELLA: a modular framework for SpatioTemporal Event-based Lagrangian particLe trAcking. Exp Fluids 67:110

\bibitem{Ni2012} Ni Z, Pacoret C, Benosman R, Ieng S, Regnier S (2012) Asynchronous event-based high speed vision for microparticle tracking. J. Microsc. 245 (3):236–244

\bibitem{Wang2020} Wang Y, Idoughi R, Heidrich W (2020) Stereo Event-Based Particle Tracking Velocimetry for 3D Fluid Flow Reconstruction. In: Vedaldi A, Bischof H, Brox T, Frahm JM (eds) Computer Vision ECCV 2020

\bibitem{Gunasekaran2025} Gunasekaran S, Mongin M, AlSattam OA and Hirakawa K, (2025) Comparison of Event-Based Alogrithms for Experimental Two-Dimensional, Two-Component Velocimetry, J. Aircraft 63:1

\bibitem{Ester1996} Ester M, Kriegel HP, Sander J, and Xu X, (1996) A Density-Based Algorithm for Discovering Clusters in Large Spatial Databases with Noise. In: Proceedings of the 2nd International Conference on Knowledge Discovery and Data Mining, Portland, OR, AAAI Press, pp. 226-231

\bibitem{Schubert2017} Schubert E, Sander J, Ester M, Kriegel HP, and Xu X (2017) DBSCAN revisited, revisited: why and how you should (still) use DBSCAN. ACM Transactions on Database Systems (TODS), 42(3), 19

\bibitem{Champagnat2014} Champagnat F, Cornic P, Cheminet A, Leclaire B, Le Besnerais G, Plyer A (2014), Tomographic PIV: particles versus blobs, Measurement Science and Technology, 25, 084002

\bibitem{Sander1998} Sander J, Ester M, Kriegel HP, and Xu X (1998), Density-based clustering in spatial databases: The algorithm GDBSCAN and its applications. Data Mining and Knowledge Discovery 2, 169–194

\bibitem{Valori2021} Valori V, Debue P, Cheminet A, Wiertel C, Padilla V, Daviaud F, Chaabo T, Ostovan Y, Cuvier C, Foucaut J-M, Laval J-P, Dubrulle B, (2021) On Lagrangian Intermittency from 4-D Particle Tracking Velocimetry measurements of a turbulent von Karman flow, Arxiv 2104.04222

\bibitem{LeBris2025} Le Bris J, Leclaire B, Cornic P, Champagnat F, Musci B, Cheminet A. (2025) Consensus based tracking for 3D-PTV at high seeding densities, Exp. Fluids 66:186

\bibitem{Manning2009} Manning CD, Raghavan P, Hinrich Schütze (2009) An Introduction to Information Retrieval, Cambridge University Press

\bibitem{Debue2021} Debue P, Valori V, Cuvier C, Daviaud F, Foucaut JM, Laval JP, Wiertel C, Padilla V, Dubrulle B (2021) Three-dimensional analysis of precursors to non-viscous dissipation in an experimental turbulent flow, J. Fluid Mech. 914:A9

\bibitem{Schanz2016} Schanz D, Gesemann S, Schröder A (2016) Shake-the-box: Lagrangian particle tracking at high particle image densities. Exp Fluids 57:04

\end{thebibliography}
\end{document}